\documentclass{emulateapj}
\usepackage{color,graphicx,lscape,natbib}
\shortauthors{Morrissey et al.}
\begin{document}
\title{The Calibration and Data Products of the Galaxy Evolution Explorer}
\slugcomment{Accepted to the Ap.J.Suppl. (special GALEX issue)}
\author{Patrick Morrissey\altaffilmark{1},
Tim Conrow\altaffilmark{1},
Tom A. Barlow\altaffilmark{1},
Todd Small\altaffilmark{1},
Mark Seibert\altaffilmark{7},
Ted K. Wyder\altaffilmark{1},
Tamas Budavari\altaffilmark{5},
Stephane Arnouts\altaffilmark{4},
Peter G. Friedman\altaffilmark{1},
Karl Forster\altaffilmark{1},
D. Christopher Martin\altaffilmark{1},
Susan G. Neff\altaffilmark{8},
David Schiminovich\altaffilmark{10},
Luciana Bianchi\altaffilmark{2},
Jose Donas\altaffilmark{4},
Timothy M. Heckman\altaffilmark{5},
Young-Wook Lee\altaffilmark{3},
Barry F. Madore\altaffilmark{7},
Bruno Milliard\altaffilmark{4},
R. Michael Rich\altaffilmark{9},
Alex S. Szalay\altaffilmark{5},
Barry Y. Welsh\altaffilmark{6}, 
Sukyoung K. Yi\altaffilmark{3}}
\altaffiltext{1}{California Institute of Technology, MC 405-47, 1200 East California Boulevard, Pasadena, CA 91125}
\altaffiltext{2}{Center for Astrophysical Sciences, The Johns Hopkins University, 3400 N. Charles St., Baltimore, MD 21218}
\altaffiltext{3}{Center for Space Astrophysics, Yonsei University, Seoul 120-749, Korea}
\altaffiltext{4}{Laboratoire d'Astrophysique de Marseille, BP 8, Traverse du Siphon, 13376 Marseille Cedex 12, France}
\altaffiltext{5}{Department of Physics and Astronomy, The Johns Hopkins University, Homewood Campus, Baltimore, MD 21218}
\altaffiltext{6}{Space Sciences Laboratory, University of California at Berkeley, 601 Campbell Hall, Berkeley, CA 94720}
\altaffiltext{7}{Observatories of the Carnegie Institution of Washington, 813 Santa Barbara St., Pasadena, CA 91101}
\altaffiltext{8}{Laboratory for Astronomy and Solar Physics, NASA Goddard Space Flight Center, Greenbelt, MD 20771}
\altaffiltext{9}{Department of Physics and Astronomy, University of California, Los Angeles, CA 90095}
\altaffiltext{10}{Department of Astronomy, Columbia University, New York, NY 10027}

\begin{abstract}
We describe the calibration status and data products pertaining to the GR2 and
GR3 data releases of the Galaxy Evolution Explorer (GALEX).  These releases have
identical pipeline calibrations that are significantly improved over the GR1
data release.
GALEX continues to survey the sky in the Far Ultraviolet (FUV,
$\sim 154$~nm) and Near Ultraviolet (NUV, $\sim 232$~nm) bands, providing
simultaneous imaging with a pair of photon counting, microchannel plate, delay
line readout detectors.  These 1.25$^{\circ}$ field-of-view detectors are
well-suited to ultraviolet observations because of their excellent red
rejection and negligible background.  A dithered mode of observing and photon
list output pose complex requirements on the data processing pipeline,
entangling detector calibrations and aspect reconstruction algorithms.  Recent
improvements have achieved photometric repeatability of 0.05 and 0.03 m$_{AB}$
in the FUV and NUV, respectively.  We have detected a long term drift of
order 1\% FUV and 6\% NUV over the mission.  Astrometric precision is of order
0.5\arcsec\ RMS in both bands.  In this paper we provide the GALEX user with a broad overview of
the calibration issues likely to be confronted in the current release.
Improvements are likely as the GALEX mission continues into an extended phase
with a healthy instrument, no consumables, and increased opportunities for
guest investigations.
\end{abstract}


\section{Introduction}

\notetoeditor{This paper is submitted for consideration in the GALEX
special issue of the Astrophysical Journal Letters.} 

We describe the calibration status and data products pertaining to the combined
GR2 and GR3 data releases of the Galaxy Evolution Explorer (GALEX).  GALEX is a
NASA Small Explorer mission performing an all-sky ultraviolet (UV)
survey in the Far UV (FUV$\sim 154$~nm) and Near UV (NUV$\sim 232$~nm) bands
with 4~--~6\arcsec\ resolution and $\sim 50$~cm$^{2}$ effective area.  At the
time of this writing, over 13500 square degrees of this survey are available to
the public via the Internet from the Multimission Archive at Space Telescope
Science Institute (MAST).\footnote{galex.stsci.edu}

The GR2 and GR3 releases have identical pipeline calibrations that are
significantly improved over the previous GR1 release.  Photometric precision at
a given epoch is 0.05 and 0.03 m$_{AB}$ in the FUV and NUV respectively.  We
have detected a long term drift of order 1\% FUV and 6\% NUV over the mission.
The astrometric precision has been improved to 0.5\arcsec\ RMS.  

\begin{figure*}
\begin{center}
\begin{tabular}{lr}
\includegraphics[width=3in]{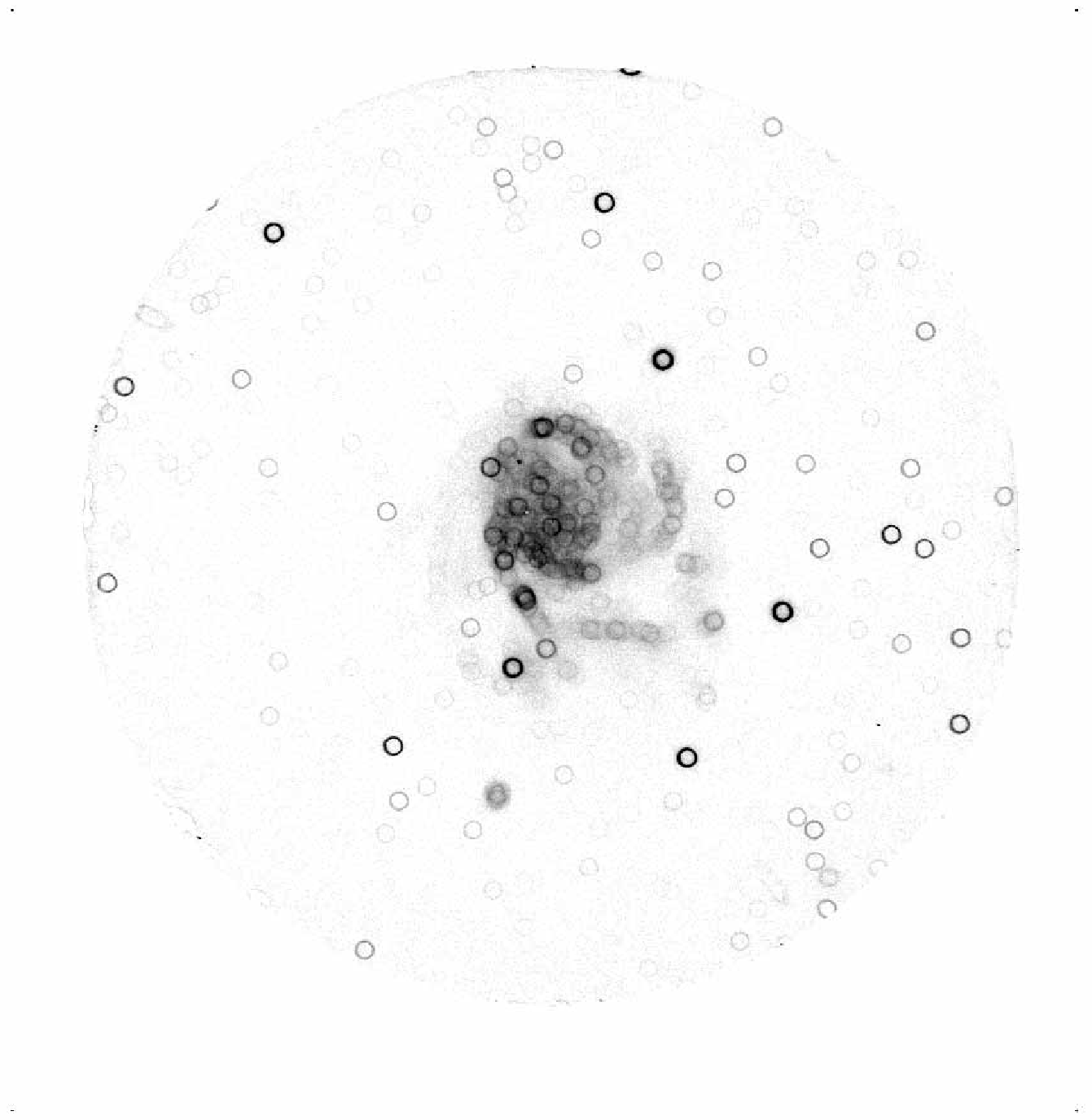} &
\includegraphics[width=3in]{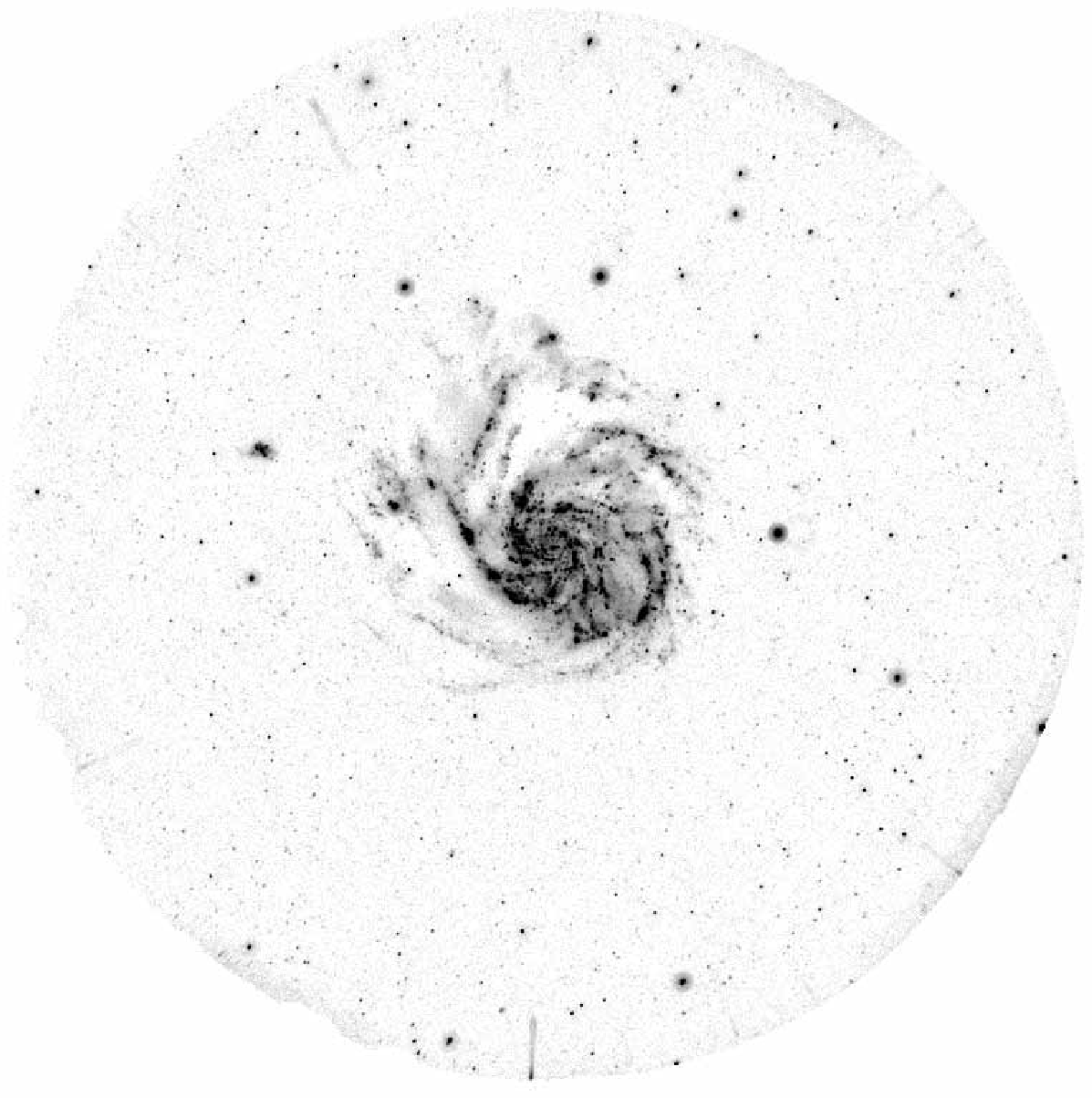}
\end{tabular}
\end{center} 
\caption{\textit{Left panel:} A GALEX NUV ``dose'' image of M101 in spacecraft
  coordinates clearly showing the 1\arcmin\ dither pattern, which is used for all
  observations to smooth out the effects of the detector flat field and reduce MCP
  fatigue. \textit{Right panel:} The fully processed, background-subtracted NUV image
  of M101 in sky coordinates, which provides the basis for the source
  catalog data.\label{dither}}
\end{figure*}

The path from raw telemetry to corrected catalogs, images and spectra is a
complicated one for the GALEX pipeline.  This fact is illustrated by
Fig.~\ref{dither}, which shows a GALEX image in detector and sky
coordinates, which are representative of the pre- and post-pipeline images.
This overview highlights the data quality,
errors and peculiarities that are likely to confront astronomers.  As is true
for any large dataset, knowledge of these basic characteristics is essential to
avoid confusion.  

The GALEX mission continues into an extended phase with a
healthy instrument, no consumables, and increased opportunities for guest
investigations.  While the survey is on-going, it is likely that we will make
improvements to the calibration and discover additional sources of error.  We
expect to provide updates to this paper on the web at
\textsc{www.galex.caltech.edu}, as well as in the literature.

This paper is organized as follows.  In Section~\ref{overview} we summarize the
design of the mission and instrument.  In Section~\ref{pipeline} we describe
how GALEX data are reconstructed and what the key pipeline products are.  In
Section~\ref{photometry} we describe the photometric calibration of the data,
including a discussion of saturation effects and drift.  In
Section~\ref{resolution} we describe the instrument resolution across the field
of view and as a function of detector gain.  In Section~\ref{astrometry}
we describe the astrometric performance of the instrument.  In
Section~\ref{spectroscopy} we describe the spectroscopy calibration and present
an example extraction.  Finally, in Section~\ref{conclusions} we summarize the
key points of the paper.

\section{Mission and Instrument Overview}
\label{overview}

GALEX was launched on an Orbital Sciences Corporation (Orbital) Pegasus rocket
on 2003 April 28 at 12:00~UT from the Kennedy Space Center into a circular,
700~km, 29\degr\ inclination orbit.  This 0.5~m diameter instrument has a very
wide 1.25$^{\circ}$ field of view sensitive to m$_{\rm AB}\sim 21$\ in the All
Sky Imaging Survey (AIS), and to m$_{\rm AB}\sim 25$\ in the Deep Imaging
Survey (DIS).  The instrument provides simultaneous FUV and NUV imaging with a
pair of photon counting, microchannel plate, delay line readout detectors.
These detectors are well-suited to ultraviolet observations because of their
excellent red rejection and negligible background.  An objective grism mode is
also available for low resolution spectroscopy.  The last comparable survey in
this wavelength range, which extended to m$_{\rm AB}\sim 9$\ was performed by
the Belgian/UK Ultraviolet Sky Survey Telescope on the TD1 satellite launched
in 1972 and described by \citet{boksenberg1973}.

GALEX makes science observations on the night side of each orbit (to keep sky
background to a minimum) during ``eclipses'' that are typically in the range of
1500~--~1800~s.  In the AIS, as many as 12 targets are observed during an
eclipse, while the Medium Imaging Survey (MIS) has one target per eclipse and
the DIS returns to the same target for 30 or more eclipses.  As of this
writing, we have observed over 19,000 square degrees of sky and accumulated
over 4 terabytes of science telemetry.  
An overview of key performance parameters is presented in Tab.~\ref{performance}.

\begin{deluxetable*}{lcc}
\tabletypesize{\scriptsize}
\tablecaption{Summary of measured performance parameters for GALEX.\label{performance}}
\tablehead{\colhead{Item} & \colhead{FUV Band} & \colhead{NUV Band}}
\tablewidth{0pt}
\startdata
Bandwidth:\tablenotemark{a}    & 1344~--~1786~\AA             
                               & 1771~--~2831~\AA \\
Effective wavelength ($\lambda_{eff}$):\tablenotemark{b}
                               & 1538.6~\AA                   & 2315.7~\AA\\
Mean effective area:           & 19.6 cm$^2$                  & 33.6 cm$^2$\\
Peak effective area:           & 36.8 cm$^2$ at 1480~\AA      & 61.7 cm$^2$ at 2200~\AA\\
Astrometry (1$\sigma$, $R \le 0.6^{\circ})$:        
                               & $\pm 0.59$\arcsec            & $\pm 0.49$\arcsec\\
Field of view:                 & 1.27$^{\circ}$               
                               & 1.25$^{\circ}$\\
Photometry (1$\sigma$):        & $\pm 0.05 m_{AB}$            & $\pm 0.03 m_{AB}$\\
Zero point ($m0_{UV}$):        & 18.82                        & 20.08\\
Image resolution (FWHM):       & 4.2\arcsec                   & 5.3\arcsec\\
Spectral resolution ($\lambda/\Delta\lambda$):
                               & 200                          & 118 \\
Spectral dispersion            & 1.64\AA-arcsec$^{-1}$       & 4.04\AA-arcsec$^{-1}$\\
Detector background (typical):\\
\hspace{2em}Total:             & 78 c-s$^{-1}$                & 193 c-s$^{-1}$\\
\hspace{2em}Diffuse:           & 0.66 c-s$^{-1}$-cm$^{-2}$    & 1.82 c-s$^{-1}$-cm$^{-2}$\\
\hspace{2em}Hotspots:          & 47 c-s$^{-1}$                & 107 c-s$^{-1}$\\
Sky background (typical):\tablenotemark{c} 
                               &  1000 c-s$^{-1}$             & 10000 c-s$^{1}$\\
Limiting magnitude ($5\sigma$):\\
\hspace{2em}AIS (100~s):       & 19.9                         & 20.8\\
\hspace{2em}MIS (1500~s):      & 22.6                         & 22.7\\
\hspace{2em}DIS (30000~s):     & 24.8                         & 24.4\\             
Linearity:\\
\hspace{2em}Global (10\% roll off):     & \multicolumn{2}{c}{18000 c-s$^{-1}$} \\
\hspace{2em}Global (50\% roll off):     & \multicolumn{2}{c}{91000 c-s$^{-1}$} \\
\hspace{2em}Local (10\% roll off):\tablenotemark{e}
                               & 114 c-s$^{-1}$               & 303 c-s$^{-1}$\\
Pipeline image format: & \multicolumn{2}{c}{$3840\times3840$ elements with 1.5\arcsec\ pixels}\\
\enddata
\tablenotetext{a}{Includes wavelengths with effective area at least 10\% of the peak value.}
\tablenotetext{b}{Wavelength-weighted: $\lambda_{eff} = \frac{\int A_{eff}\times \lambda}{\int A_{eff}}$}
\tablenotetext{c}{These correspond to 246 and 601~photons-s$^{-1}$-cm$^{-2}$-sr$^{-1}$-\AA$^{-1}$, respectively.}
\tablenotetext{d}{These are worst-case values for point sources.}
\end{deluxetable*}

\begin{figure}
\begin{center}
\includegraphics[width=3in,angle=0]{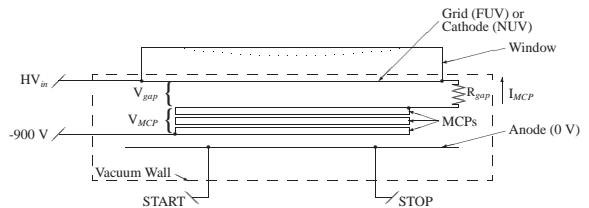}
\end{center}
\caption{GALEX sealed tube detector head electro-mechanical block diagram.  The
NUV and FUV tube heads are nearly identical with the principal differences
being the choice of photocathode material (CsI for FUV and Cs$_2$Te for NUV)
and window material (MgF$_2$ for FUV, SiO$_2$ for NUV).  Also, the cathode
material is deposited on the detector window in the NUV channel but directly on
the MCP in the FUV channel.  Instead of a cathode, the FUV window has a charged
grid of wires that enhances the sensitivity of the detector.  The NUV cathode is
proximity focused on the front MCP, and due to its reentrant design is more
prone to reflections from off-field sources.  $HV_{in}$ represents the negative
applied voltage, 5-6~keV, while a resistor chain ($R_{gap}\sim 10$~MOhm in
series with the MCPs at about $30$~MOhm each) sets the voltage across the gap
($V_{gap}\sim 500$~V) and MCP stack ($V_{MCP}\sim 4500$~V) driving a detector
bias current $I_{MCP}\sim 50\ \mu$A.  The detector pulse outputs in one
dimension are labeled START and STOP (or sometimes ``A'' and
``C'').\label{detector_head}}
\end{figure}

A GALEX detector head is shown schematically in Fig.~\ref{detector_head}.
Individual photons incident at the cathode set off an isolated avalanche of
current inside parallel 10-12~micron diameter microchannel plate (MCP) pores as they accelerate through a
few-keV field.  The resulting charge cloud at the rear of the MCP stack,
approximately $10^7$ electrons in size, is deposited on a double-layer anode,
divided, and finally measured at each of four outputs.  The layers of the anode
form a pair of orthogonal delay lines.  Timing differences measured for the
charge pulses at each corner of the anode are proportional to the position of
the initiating photon event.  For faint UV observations, the combination of low
background (zero read noise) and high red rejection that these detectors
possess represents a favorable trade-off against the superior QE and field
flatness of more conventional CCD detectors.  Furthermore, MCP detectors do not
require cooling, an important consideration in the contamination-sensitive UV
band.  In order to improve photometric performance, observations are dithered
in a spiral pattern approximately 1\arcmin\ in diameter (at a rate of
approximately 0.5 revolutions per minute) that smooth out small scale
variations in the flat field and reduce bright source fatigue on the MCPs.

A more detailed instrument overview is provided in
\citet{morrissey2005}, while the mission design is described by
\citet{martin2005}.  Further details of our mission operations experience are
described in \citet{forster2006} and \citet{morrissey2006}.

\section{Image Reconstruction Overview}
\label{pipeline}
The GALEX detector output is a list of time-ordered 40-bit photon positions and
pulse heights.  These data are delivered to the instrument Data Processing Unit
(DPU), which adds a timing stamp every 5~ms (or every 20~$\mu$s in uncompressed
mode).  The time stamp is derived from the spacecraft clock, which has a drift
of the order of 1~s each week that is corrected with ground updates.  Data are
stored in the 2.8~GB ring buffer of a Solid State Recorder (SSR), which is
down-linked to a ground station approximately four times each day.

The ground pipeline reconstructs a list of photon positions and arrival times
into an image.  Broadly speaking, the photon lists are corrected for detector
effects and image motion, projected into sky coordinates, binned into images,
flat fielded, and finally background subtracted.  These final image products
are then scanned by the open-source SExtractor program \citep{bertin1995} to
produce catalogs of source positions and fluxes.  The steps along this path are
described in more detail in the following sections.

\subsection{Static Calibration}

The static calibrations are those used to assemble raw detector event positions
into accurate locations in the GALEX focal plane.  They also flag events that are not
photon generated, such as the periodic ''STIM'' pulses that are always
electronically superimposed (in parallel with real photon events) on the
detector anodes, and regions on the MCPs that are dominated by hotspots.  

Each photon event in the GALEX science telemetry is a heavily encoded group of
fields that, when decoded and assembled, produces key values listed in
Tab.~\ref{40bit}.  In the GALEX design \citep{jelinsky2003}, a free-running
coarse clock subdivides the large detector areas into an $8\times 8$ grid.
Detector event pulses are measured asynchronously while this clock runs, so
that each position (proportional to the measured arrival time difference of
pulses from either side of the delay line anode on each axis) comprises an
integer number of coarse clock steps plus a fine position (essentially the
coarse clock phase).
\begin{deluxetable}{lll}
\tablecaption{Key data contained in the GALEX RAW6 format photon lists.
  Photons from the detector itself are not time stamped but rather time
  ordered, and the time stamp is added by the DPU
  before being written to the SSR.\label{40bit}}
\tablehead{\colhead{Column} & \colhead{Bits} & \colhead{Description}}
\startdata
t           & 40   & Event detection time\tablenotemark{a}\\
$X_{AmC}$   & 12   & Raw X detector position in FEE pixels \\
$Y_{AmC}$   & 12   & Raw Y detector position in FEE pixels \\
$X_A$       &  5   & Wiggle (the phase of the TAC)\\
$X_B$       &  3   & X axis coarse clock\\
$Y_B$       &  3   & Y axis coarse clock\\
Q           &  5   & Detected pulse height \\
\enddata
\tablenotetext{a}{In high resolution mode, the timing precision uses 40
  bits to attain 20~$\mu$s resolution, however this mode is rarely used.  For
  most data, only 32 bits are significant, attaining 5~ms resolution.}
\end{deluxetable}

In order to construct a raw source position $(X,Y)$ from these data, one would apply the
following formulas: 
\begin{eqnarray}
X & = & X_{AmC} + \alpha X_B \\
Y & = & Y_{AmC} + \beta  Y_B,
\end{eqnarray}
where the constants $\alpha$ and $\beta$ are in the range of 2000 (solved by
minimizing source widths across the field of view), $X_B$ and $Y_B$ are the
coarse clock counters, and $X_{AmC}$ and $Y_{AmC}$ are the ''fine'' position
measurements representing the coarse clock phase.\footnote{The notation derives
from the ends of the delay line ``A'' and ``C'', or sometimes ``START'' and
``STOP'' in other work, thus the position is proportional to a timing difference of $A-C$ or
``AmC.''} After the detector positions are assembled, the
following static corrections are applied (in order) to correct the raw
positions to a linear representation of the actual source position in the focal
plane:
\begin{itemize}
\item{\textbf{Centering and scaling:} Coarsely scales the raw positions from bins to
   microns.  The scale factors are different for each axis.}
\item{\textbf{Wiggle:} The GALEX coarse-bit sub-array approach allows
  degenerate combinations of coarse and fine components for a given detector
  location.  Each of these may have a slightly different electronic
  non-linearity correction owing to the differing phases of the
  time-to-amplitude converter (TAC) for each event. The image of a single point
  source on the detector is thus \textit{blurred} by the superposition of all
  of the TAC data, an effect referred to as wiggle.  This contrasts with
  smaller, more conventional time-to-amplitude conversion systems. These do not
  utilize a coarse counter bit and translate all electronic
  non-linearities into source \textit{position} errors rather than source
  \textit{widths}.  With careful part selection, ``wiggle'' in the flight
  units has been minimized considerably.  In order to eliminate it completely,
  each detected event includes a 5-bit value $X_A$, which represents where in
  the TAC range each measurement was made. Residual electronic
  non-linearities are then removed using lookup tables for combinations of
  $(X,X_A)$ and $(Y,X_A)$.  Both axes use the same $X_A$ data because
  the $X$ and $Y$ pulses are correlated in time for a given detector
  position.}
 \item{\textbf{Walk:} This transformation corrects position-vs-pulse-height (Q) errors.  A lookup table correction applied in a similar manner as
  wiggle, reflective of the
  fact that the electronics can not perfectly compensate for pulse sizes that
  vary by a factor of 10 across the detector.}
\item{\textbf{Spatial non-linearity (distortion):} A lookup table corrects
  high voltage detector edge effects, which cause electrons
  to bend from their optimal path, distorting the image especially near the
  edge of the field.}
\item{\textbf{Hotspot masking:} Each detector has numerous small, high count
  rate features resulting from a combination of microscopic MCP debris and
  channel defects that combined generate about half of the intrinsic detector
  background.  These spots come and go, and are best handled by
  direct masking, which eliminates only about a percent of the detector area.
  New masks are generated roughly annually; hotspots are generally easy to pick
  out of images because they ``reverse dither'' (since they are fixed in
  detector space) and create $\sim1$\arcmin\ circular artifacts in GALEX
  sky-space images.  In the data pipeline, the hot spot masks create
  localized regions of zero effective sensitivity in detector space, one reason
  why dithered observations are preferred for these detectors.  Depending
  on pointing accuracy in a given observation, the dither typically maps these
  regions into ''donuts'' of reduced (but not zero) response on the sky.}
\end{itemize}

\subsection{Dynamic Aspect Correction}

The fully-corrected event positions in spacecraft coordinates accurately record
where each photon landed in the GALEX focal plane.  The next step is to
reconstruct the spacecraft pointing on the sky as a function of time in order
to remove the dither pattern from the observed data.  This is the 3-parameter
``aspect,'' which includes the right ascension, declination, and roll of the
telescope boresight.  For imaging mode data, a precise estimate is arrived at
by refining the spacecraft Attitude Control System (ACS) pointing estimate with
NUV data\footnote{The process will run with FUV-only data, although owing to
the generally decreased numbers and brightnesses of stars in FUV data, some
parameters such as the interpolation interval need to be adjusted.} binned into
1~s slices.  These snapshot images are each matched to the ACT catalog
\citep{urban1998} of known stars,\footnote{Occasionally, matches to known stars
cannot be found, usually due to pointing errors.  In these rare cases, a good
Point Spread Function (PSF) is achieved by performing the aspect refinement
relative to \textit{any} sources of emission detected in the field that are
reasonably compact. This is called relative aspect refinement mode and results
in images with good PSFs but poor absolute astrometry. Roll is not refined in
relative mode.}  and the average error vector from the matches to the catalog
stars are used to establish a correction to apply to the ACS estimate.  The
source positions are then transformed to sky coordinates (J2000) using a
Gnomonic projection based on the pointing at the interpolated time.

Grism data undergoes a similar process, but the software can not match spectra
to known stars at this stage of processing.  A relative aspect solution is
performed based solely on the instantaneous spectral image positions, and
astrometric refinement is added at a later stage.

Once sky coordinates have been assigned to each photon, an "extended photon
list" is computed for internal use by the pipeline in forming the final data
products.  In cases where detailed, time-dependent behavior of one or more
objects over the course of an exposure is of interest, the extended photon list
can be written out to a so called "x-file," by special request.

\subsection{Pipeline Products}

Images and catalogs from various stages of pipeline processing reveal many
characteristics of the complex, multidimensional GALEX dataset and are included
with each data release as pipeline products.  A description of key products is
provided here for reference.

The corrected photon data are binned into images and flux calibrated in order
to form the basis for the GALEX catalogs.  Count images (\textit{-cnt}), which
contain aspect-corrected detector counts per sky pixel are scaled by the
effective area and exposure time to form the flux-calibrated intensity
(\textit{-int}) images.  This step is accomplished with the relative response
(\textit{-rr}) image, which contains the effective exposure for each pixel:

\[rr(X,Y) = \Sigma_{t=0}^{T} flat(X(t),Y(t)) \times (1 - f_{dead}(t)) \times dt(t),\]

where $flat(X(t),Y(t))$ is the detector flat field\footnote{The flat field has
  a value of 1 in regions for which a zero-point-magnitude source generates
  exactly 1~deadtime-corrected photon-s$^{-1}$.} shifted to properly align with
  the dithered image time slice, $f_{dead}$ is the global electronic dead time
  based on the measured STIM pulser rate during each time slice,\footnote{Since
  the electronic STIM pulses occur at a precisely known rate, the fraction that
  are missing from the event list over a given interval will be known. This is
  the fractional dead time, which corrects the global rates only and does not
  account for local MCP saturation.} and $dt(t)$ is the fraction of each 1~s
  time slice during which data was collected (i.e. HV was on and the detector
  was otherwise in a nominal observing mode).  Building the effective exposure
  time in this way properly handles the fact that the edges of the image
  receive less exposure than the middle, and that the sources each sample a
  1\arcmin\ region on the flat field.  The high resolution relative response
  image (\textit{-rrhr}), which properly matches the count and intensity map
  pixel scales, is up-sampled from the 6\arcsec\ relative response array.  The
  intensity image is simply the count image divided by the high resolution
  relative response image.

As a final step, the background in each intensity image must be estimated and
removed.  Since the GALEX detectors have negligible intrinsic background and
the UV sky is dark, estimation of the true background is not properly handled
by the code within SExtractor.  Instead, we create a custom threshold image for
SExtractor that allows a bypass of the normal process for distinguishing low
rate sources from background.  The typical sky background count rates in high
Galactic latitude GALEX pointings are $\sim 10^{-4}$ ($\sim 10^{-3}$)
photons-s$^{-1}$-arcsec$^{-2}$ in the FUV (NUV), thus it is necessary to use
the full Poisson distribution.  The background is determined in a set of large
bins, 192\arcsec\ on a side. In images with exposure times greater than
10,000~s, the bins are 96\arcsec\ on a side. For the Poisson
distribution, the probability of observing greater than or equal to $k$ events
for a mean rate $x$ is given by the incomplete gamma function $P_k(x)$
\citep{press1992}.  The average is computed iteratively within each background
bin by clipping out high pixels where $P_k(x) < 1.35 \times 10^{-3}$, a
probability equivalent to $3\sigma$ for a Gaussian distribution. Even after
clipping, bright or extended sources can still bias the background
determination, therefore, a $5\times5$ median filter is applied to the
background bins. The resulting background values are linearly interpolated to
the full resolution of the images.  We have determined that the background may
be biased by a few percent due to contamination from the wings of objects. To
address this problem, background estimation and source extraction software is
run twice on each GALEX image. The second run uses source position data from
the first iteration to mask out source pixels and recompute the sky background
and background-subtracted intensity images.  The resulting background image
(\textit{-skybg}) is subtracted from the intensity image to form the
background-subtracted intensity image (\textit{-intbgsub}), which is the image
scanned by SExtractor.

There remain issues with the background at the 5\% level that affect the MIS
survey at the faint end.  Furthermore, there are numerous artifacts not removed
by the current pipeline that the user should be alert for.  These include:
bright star glints in the NUV images near the edge of the detector window;
large out-of-focus pupil images resulting from reflections in the imaging
window and dichroic beam splitter; a 1\arcmin\ diameter skirt around bright NUV
sources (at the few-percent level in total energy) resulting from the
proximity-focused photocathode; and photoemission and diffraction from the
quantum-efficiency (QE) enhancing grid wires on the FUV detector window that
appear as linear features around bright sources.  Work is being done to flag
and remove these features automatically in the pipeline software, but at
present they amount to a small fraction of the survey coverage area ($\sim
1$\%).

The primary pipeline image products and descriptions are listed in
Tab.~\ref{arrayproducts}, while catalog products are listed in
Tab.~\ref{tableproducts}.  All of the catalog output products are generated by
SExtractor, with the exception of the very useful merged catalog
(\textit{-mcat}), which combines FUV and NUV data in one table and adds derived
columns including the detector coordinates of a source, detector intrinsic
pulse height, number and brightness of nearby neighbors to a source, and many
more.  Matches between the bands are made by assigning a match probability to
each pair of sources using the source separations and positional uncertainties.
Sources with signal-to-noise (S/N) ratio of less than 2 or a separation greater
than 3\arcsec\ are not matched across bands (but do appear in the merged
catalog).  GALEX fields are uncrowded enough that the probability of a
mis-match is very small, however a multiple candidate chain technique is used
to assure that candidates are always matched the same way regardless of the
order in which the process is started.

\begin{deluxetable*}{lll}
\tablecaption{Binary tables generated by the GALEX pipeline.\label{tableproducts}}
\tablehead{\colhead{Extension} & \colhead{Description}}
\startdata
-cat       & Primary SExtractor source catalogs for each band\\
-fcat      & SExtractor catalog of NUV extractions based on FUV coordinates\\
-gauss     & Listing of the Gaussian filter elements used by SExtractor for image convolution\\ 
-mcat      & Merged catalog for both NUV and FUV extractions\\
-ncat      & SExtractor catalog of FUV extractions based on NUV coordinates\\
-sexcols   & Listing of SExtractor catalog column names\\
-sexparams & Listing of SExtractor parameters used\\
\enddata
\end{deluxetable*}

The GALEX data products are made available to the general public through the
GALEX Science Database at MAST.  A dedicated website provides access to the
pipeline output files via a web-based browser, as well as other derived
products, such as JPEG images. The catalog data are also stored in a commercial
SQL Server database engine fine-tuned for astronomy applications, which
provides sophisticated query interfaces supporting, for example, efficient
spatial and color searches.  The GALEX MAST database also contains enhancements
over the photometric pipeline outputs, e.g.\ spatial crossmatching to the Sloan
Digital Sky Survey (SDSS) catalog. Also, since adjacent GALEX fields overlap
slightly, a \textit{primary resolution} algorithm deals with the problem of
unique sample selection.  The formal GALEX footprint definitions of the GALEX
surveys are published in the National Virtual Observatory's online Footprint
Service \citep{budavari2006}.  On this
site,\footnote{http://voservices.net/footprint} one can not only search and
visualize the GALEX sky coverage but also compute the exact area in square
degrees or intersect it with other surveys.

\section{Photometry}
\label{photometry}

\subsection{GALEX Magnitude Reference System}
\label{magref}
The GALEX photometric calibration is tied to the \textit{Hubble Space
Telescope} (HST) through reference spectra supplied in the CALSPEC database and
described by \citet{bohlin2001}.  Almost all of the white dwarfs in this
database are too bright for a photon counting instrument such as GALEX in
direct imaging mode.  As such, we are restricted to the dimmest sources in the
catalog, which are all ultimately based on the \textit{International
Ultraviolet Observer} (IUE) calibration.  All of these data have been
reprocessed for the HST calibration program, as detailed in \citet{bohlin1996}.
The GALEX calibration is thus traceable to a hydrogen white dwarf model scaled
using the ground based photometry of \citep{landolt1992} in combination with
several white dwarf transfer standards measured with the HST Faint Object
Spectrograph.  The combination of models and measurements was used to define
corrections to the IUE calibration, thus transferring the HST calibration and
allowing a fairly large group of standards from the IUE library to enter the
CALSPEC database.  Of these, we have observed 18 as detailed in
Tab.~\ref{standards}, however we use one of the dimmest, LDS749b, as our
primary standard for direct imaging.  For the most part this white dwarf is in
the linear range of the GALEX detectors -- bright enough for high
S/N measurements in just a few minutes, but dim enough not to
saturate the microchannel plates.  At the edge of the field in the NUV, where
the detector gain is lowest, we apply a $\sim 20$\% correction for local
saturation.  This effect was discovered relatively recently as our database of
white dwarf observations has expanded to include a much finer sampling of the
detectors in regions where the gain is an issue.\footnote{Since the detector
MCPs are clamped at the edges, there is less chance for a charge packet at the
edge of the field to spread laterally while traversing the stack.  At the
detector center, where there are small gaps between the plates, charge can
spread across the pores.  Thus the edges of the detectors tend to saturate more
quickly than the centers.  This effect can be seen for bright objects with
multiple measurements by plotting the observed magnitude against the
unsaturated detector pulse height, \textsc{\#uv\_q}, tabulated in the merged
catalog for each source.  This quantity is the \textit{normal} unsaturated
pulse height of the detector at the location the source was measured.  Note
that the effect is minimized for large measuring apertures because the core of
the PSF is primarily affected.}  At the time of this writing we are planning
observations of the (dimmest) CALSPEC white dwarf, LB227, in order to improve
our confidence in the NUV zero point.  Unfortunately this source is not much
dimmer than LDS749b, but since we believe the saturation of LDS749b is mild it
should provide very useful verification of our saturation correction.

\begin{deluxetable*}{lllll}
\tablecaption{\textit{HST} White Dwarf Standards observed by GALEX.\label{standards}}
\tablewidth{0pt}
\tablehead{
\colhead{Star} 
& \colhead{m$_{\rm FUV}$\tablenotemark{a}} 
& \colhead{m$_{\rm NUV}$\tablenotemark{a}} 
& \colhead{$\alpha$(2000)\tablenotemark{b}} 
& \colhead{$\delta$(2000)\tablenotemark{b}}}
\startdata
GD50        & 11.98 & 12.57 & \phantom{0}3$^h$ 48$^m$ 50.4$^s$ 
            & \phantom{0}-1$^{\circ}$ \phantom{0}1\arcmin\ 26.8\arcsec\\ 
HZ4         & 14.53 & 14.50 & \phantom{0}3$^h$ 55$^m$ 22.0$^s$ 
            & \phantom{-0}9$^{\circ}$ 47\arcmin\ 18.2\arcsec\\
HZ2         & 12.86 & 13.25 & \phantom{0}4$^h$ 12$^m$ 43.6$^s$ 
            & \phantom{-}11$^{\circ}$ 51\arcmin\ 49.9\arcsec\\
G191B2B     & -     & 10.17 & \phantom{0}5$^h$ \phantom{0}5$^m$ 30.7$^s$ 
            & \phantom{-}52$^{\circ}$ 49\arcmin\ 49.3\arcsec\\
GD108       & 12.39 & 12.77 & 10$^h$ \phantom{0}0$^m$ 47.3$^s$ 
            & \phantom{0}-8$^{\circ}$ 26\arcmin\ 28.5\arcsec\\
HZ21        & 12.55 & 13.13 & 12$^h$ 13$^m$ 56.3$^s$ 
            & \phantom{-}32$^{\circ}$ 56\arcmin\ 30.8\arcsec\\
GD153       & 11.33 & 11.91 & 12$^h$ 57$^m$ \phantom{0}2.4$^s$ 
            & \phantom{-}22$^{\circ}$ \phantom{0}1\arcmin\ 53.6\arcsec\\ 
HZ43        & 10.75 & 11.36 & 13$^h$ 16$^m$ 21.8$^s$ 
            & \phantom{-}29$^{\circ}$ \phantom{0}5\arcmin\ 57.8\arcsec\\
HZ44        & 10.02 & 10.27 & 13$^h$ 23$^m$ 35.0$^s$ 
            & \phantom{-}36$^{\circ}$ \phantom{0}7\arcmin\ 58.8\arcsec\\
GRWp70d5824 & 12.14 & 12.75 & 13$^h$ 38$^m$ 49.2$^s$ 
            & \phantom{-}70$^{\circ}$ 17\arcmin\ \phantom{0}4.6\arcsec\\
SAP041C     & -     & 16.69 & 14$^h$ 51$^m$ 58$^s$ 
            & \phantom{-}71$^{\circ}$ 43\arcmin\ 17.4\arcsec\\
BD$+33^{\circ}2642$ & 10.51 & 10.47 & 15$^h$ 51$^m$ 59.9$^s$ 
            & \phantom{-}32$^{\circ}$ 56\arcmin\ 54.8\arcsec\\
SAP177D     & -     & 18.35 & 15$^h$ 59$^m$ 13.5$^s$ 
            & \phantom{-}47$^{\circ}$ 36\arcmin\ 41.4\arcsec\\ 
SAP330E     & -     & 17.84 & 16$^h$ 31$^m$ 33.8$^s$ 
            & \phantom{-}30$^{\circ}$ \phantom{0}8\arcmin\ 46.3\arcsec\\ 
LDS749B\tablenotemark{c}     
            & 15.57 & 14.71 & 21$^h$ 32$^m$ 16.4$^s$  
            & \phantom{-0}0$^{\circ}$ 15\arcmin\ 15.0\arcsec\\
G93-48      & 12.14 & 12.39 & 21$^h$ 52$^m$ 25.2$^s$  
            & \phantom{-0}2$^{\circ}$ 23\arcmin\ 17.9\arcsec\\

NGC7293     & 10.93 & 11.70 & 22$^h$ 29$^m$ 38.5$^s$  
            & -21$^{\circ}$ \phantom{0}9\arcmin\ 47.9\arcsec\\

LTT9491     & 16.09 & 14.58 & 23$^h$ 19$^m$ 35.4$^s$  
            & -18$^{\circ}$ 54\arcmin\ 32.0\arcsec\\
\enddata
\tablenotetext{a}{Magnitudes shown are predictions based on the GALEX bandpass
and publicly available reference spectra from the \textit{HST} CALSPEC database
            at http://www.stsci.edu/instruments/observatory/cds/calspec.html.}
\tablenotetext{b}{Coordinates are GALEX-measured in the NUV. 
One exception is BD$+33^{\circ}2642$, which was only observed in grism mode; the coordinate provided is 
from \citet{bohlin2001}.}
\tablenotetext{c}{This white dwarf is the primary GALEX standard.}
\end{deluxetable*}

\subsection{Magnitude Zero Point}
\label{drift}
GALEX uses the AB magnitude system of \citet{oke1983} with the FUV and NUV
magnitudes defined as follows:
\[m_{UV} = m0_{UV} - 2.5\log f_{UV}\]
Here, $f_{UV}$ is the dead-time-corrected count rate for a given source
\textit{divided by the flat field map,} which has a median value of order 0.85,
and $m0_{UV}$ is the zero point that corresponds to the AB magnitude of a
1~count-s$^{-1}$ (cps) flat-field-corrected detection.  There are independent
relative response functions, and the zero points are:
\begin{eqnarray}
m0_{FUV} & = & 18.82 \\
m0_{NUV} & = & 20.08. 
\end{eqnarray}
The GALEX project has maintained these zero points constant; they are based on
the imaging-mode bandpass determined during ground calibration.  Improvements
in calibration have been incorporated by adjusting the flat field, which is a
stacked sky background image modified by an interpolated grid of white dwarf measurements.
Since GALEX has a large field of view, this has the effect of moving the
positions in the field of view for which sources with zero point magnitudes
correspond to a true 1~count-s$^{-1}$ at the detector.  For the most part this
is not a major consequence since the flat field peak-to-valley variations for
most of the useful field amount to about a factor of 2 even in FUV, and just
15\% in NUV.  There is thus only a modest effect on the estimation of the true
expected noise for a given source with field position.  In order to estimate
any systematics resulting from application of the new flat field, a large
sample of sources common to both the GR1 and GR2/GR3 releases have been
compared.  We find that \textit{on average} sources in GR2/GR3 are shifted 0.04
and 0.01 $m_{AB}$ fainter in FUV and NUV, respectively, compared to GR1.  One
must of course be careful applying these figures to any individual measurement
because the flat has changed shape as well as its average value.

\begin{figure}
\includegraphics[width=2.25in,angle=90]{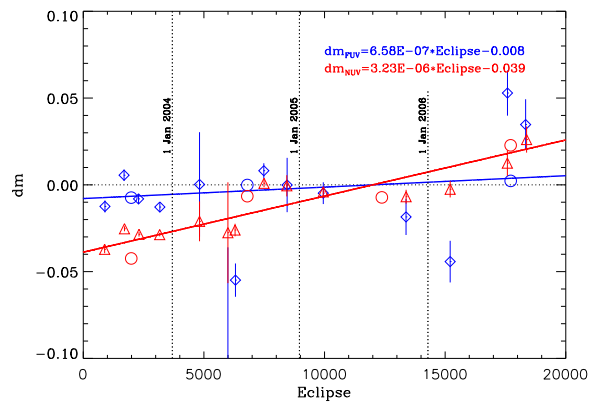}
\caption{The measured photometric drift for deep field sources between
  magnitudes 15 and 20 over the course of the GALEX mission.  FUV data is shown
  in blue: diamonds for the deep field sources and circles for LDS749b.  NUV
  data is shown in red: triangles for the deep field sources and circles for
  LDS749b.  The deep field data shown represent the mean binned in 1000 eclipse segments.  
  The drift is defined to
  be zero at eclipse 12000, which occurred in summer 2005 during the extensive
  white dwarf observing campaign upon which the current calibration is mostly
  based.  It is not clear if the trend is due to contamination in the
  instrument, or due to a mild global gain decrease (fatigue) in the
  detector microchannel plates.\label{contamination}}
\end{figure}

We have measured a drift in GR2/GR3 photometry of $0.003\pm
0.001$~mag-year$^{-1}$ FUV and $0.016\pm 0.0005$~mag-year$^{-1}$ NUV.  This was
identified with two independent means.  In the first method (yielding the
quoted results), we compared the GR2/GR3 photometry of 197 FUV and 547 NUV deep
field sources between magnitudes 15 and 20 measured at approximately yearly
intervals over the course of the mission.  We have checked this result against
the LDS749b white dwarf data, however this second result required considerable
reprocessing.  By design, the GR2/GR3 sample of white dwarf measurements
through 2005 shows no drift.  Each
measurement was given equal footing in order to improve the
spatial sampling, which only included 20 measurements per year early in the mission.  
In 2005, we identified the
NUV white dwarf saturation effect in low gain areas, and also recovered the FUV
detector from a major anomaly.  In light of these circumstances, we implemented
an extensive recalibration campaign.  Over 1500 NUV measurements were added,
but only 137 in the FUV because the detector was still completing its recovery.
Thus the GR2/GR3 flat, generated with all of the white dwarf samples from the
mission through 2005, is biased to 2005 in both bands and very strongly so in
NUV.  However, areas of the flat field sampled only in, say, 2003, will reflect
the proper calibration for that epoch, which means that there is an additional
systematic built into the flats for isolated areas.  To address the drift, we reprocessed all of the GALEX LDS749b white
dwarf data using a new flat field based on a grid of 165 LDS749b field
measurements made over 2 days in 2006.  We used a nearby field star of moderate
NUV count rate rather than the bright LDS749b to generate the NUV flat (for
FUV we used the white dwarf), mitigating concerns about the NUV
saturation.  The results of the white dwarf and deep field analyses are
consistent with each other and are presented together in Fig.~\ref{contamination}.
The data are consistent with a linear drift (although less certain
in the FUV band). We define the time of zero correction to be at eclipse
12000 based on the weight of that epoch in generating the GR2/GR3 flats.  

We have not included an FUV LDS749b data point for 2005
in Fig.~\ref{contamination}.  This is because of a second issue we discovered
during this analysis, which is that the 2005 FUV white dwarf data reads bright
by about 5\% when reprocessed with the 2006 single-epoch flat field.  This is a result of the way the measurements were made in
2005 while the FUV detector was completing its recovery from an anomaly.  During
that period (approximately a month), the detector was being cycled on and off
for only very short observation periods ($\sim$minutes) while its stability was
being tested.  Since the detector was ramped to high voltage literally minutes
before the observation began, it was still warming up during the observations,
offsetting the calibration slightly.  This means that there is a
field-dependent systematic built into the GR2/GR3 FUV flat field of
approximately 5\% in the sense that most (about 77\%) objects are actually
brighter in the FUV than they appear.  NUV is unaffected.

We have not determined if the observed drift is due to contamination or a global MCP fatigue.  
Since the performance of each MCP pore is sensitive to its total exposure,
We would expect fatigue to affect NUV more since it typically 
detects
photons at about 10 times the FUV rate.  Furthermore, contamination would normally be expected to preferentially affect the FUV band.  On the other hand, the effect we observe appears
to affect bright and dim objects the same.
Whichever effect is at play, we expect to apply
a correction automatically in the pipeline during a future calibration
update, however it is not accounted for in GR2/GR3.

\subsection{Relative precision}

\begin{figure}
\includegraphics[width=2.25in,angle=90]{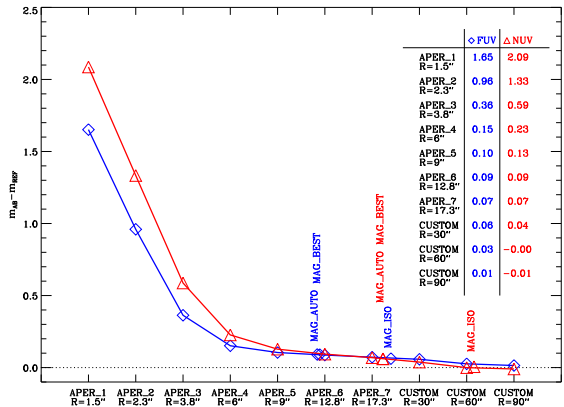}
\caption{GALEX photometric ``curve of growth,'' which shows the median
magnitude offsets relative to the largest aperture used for the white dwarf
LDS749b.  NUV measurements have been restricted to high pulse height areas of
the detector to minimize any effect due to saturation.\label{growthcurve}}
\end{figure}

For accurate and repeatable measurements of unresolved sources, aperture
magnitudes are usually best since they are fixed from measurement to
measurement.  One needs to take care to choose an appropriate aperture for
the task at hand.  Smaller apertures are more useful for unresolved sources,
but larger ones become desirable, even for point sources, when comparing the photometry of objects
measured at the center and edge of the field (where the PSF is degraded), or
for bright objects that have saturated cores.  For the purposes of this paper,
the SExtractor \textsc{mag\_aper\_4} magnitude is used unless otherwise noted.
This 12\arcsec\ diameter aperture is a good compromise between the very large
apertures required to gather almost all of the GALEX flux in the extended PSF
(as is done for the white dwarf calibration) and the small ones that are
optimal for minimizing the contribution of background.  The relationship
between the various standard SExtractor aperture magnitudes and the large
tailor-made apertures used for white dwarf calibration are presented in
Fig.~\ref{growthcurve}.

In the GR2/GR3 release, the scatter in white dwarf measurements has been
reduced to about 0.02~mag RMS in NUV and 0.045~mag RMS in FUV.  This compares
to a scatter of order 0.1~mag in both bands for the GR1 release.  We attribute
the improvement in precision to the way the flat field has
been generated, using white dwarf measurements to modify the flat locally
rather than using the average to scale a sky background field.  As with any
instrument, care should be taken when interpreting the magnitudes of
objects expected to have significantly biased spectral content compared to the
white dwarf.  These may have more scatter due to optical effects across the
field of view.

We have also verified the performance of the flat field by means independent of
the partially saturated (in the NUV) white dwarf standard.  For this purpose,
we have examined the sample of data from the DIS described in
Section~\ref{drift}.  This comprises a series of fields with several dozen
full-eclipse-length observations, many at different roll angles (thus sampling
multiple detector locations with the same source).  By matching compact ($\le
10$\arcsec\ FWHM) sources with no neighbors closer than 12\arcsec\ in the
co-added catalogs to the individual visit catalogs, a "visit" database of
sources and magnitude errors (with the co-added catalog considered to be the
truth catalog) has been generated.  A comparison of high and low gain detector
measurements for a given source reveals the count rate required to cause local
saturation.  For both detectors, the onset of saturation is at approximately
m$_{AB}\sim15$ (corresponding to 34~counts-s$^{-1}$ FUV and 108~counts-s$^{-1}$
NUV) \textit{in the lowest gain regions of the detector}.  We thus restrict
further analysis in this Section to the range $15 \le m_{AB} \le 17$.

\begin{figure}
\includegraphics[width=3.5in]{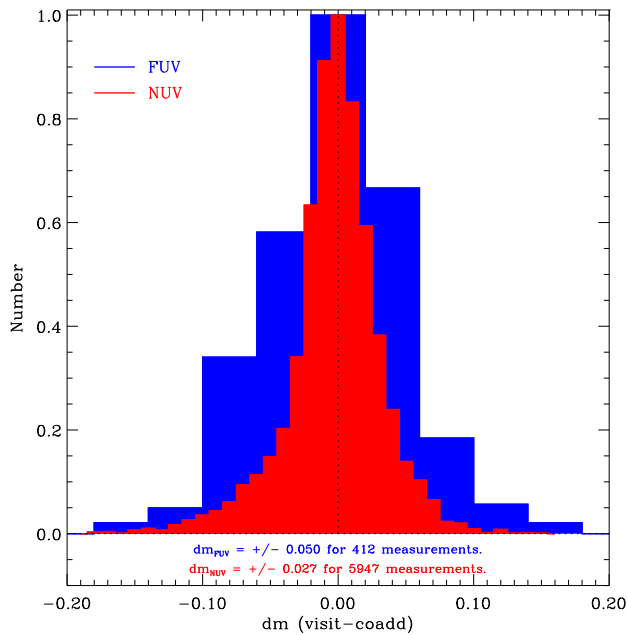}
\caption{GALEX photometric precision as determined by repeated measurements of
  bright but unsaturated sources in the GALEX deep fields.  These measurements
  are in good agreement with the results determined directly from repeat
  measurements of the white dwarf calibrators.\label{dmhist}}
\end{figure}

With the unsaturated magnitude range determined over the whole field, we can
proceed to compute the photometric repeatability as shown in
Fig.~\ref{dmhist}.  Here, we present histograms of the 12\arcsec-diameter
aperture magnitude errors in each band.  The Poisson error associated with each
detection is of order 0.5\% to 1.5\%, while the total scatter we measure in the
bright deep field sources is $\pm 0.05\ m_{AB}$ RMS in FUV and $\pm 0.03\
m_{AB}$ RMS in NUV.  These figures agree well with the white dwarf results,
thus concerns about the (mild) LDS749b NUV saturation correction are reduced.
If the white dwarf saturation had not been properly accounted for, the flat field
would have been adversely affected in the NUV low gain areas, which would in turn
increase the photometric scatter of dimmer (definitely unsaturated) deep field
sources in those same regions, and thus we would not expect good agreement in
the scatter between the white dwarf and deep field samples.

\begin{figure*}
\includegraphics[width=3in,angle=90]{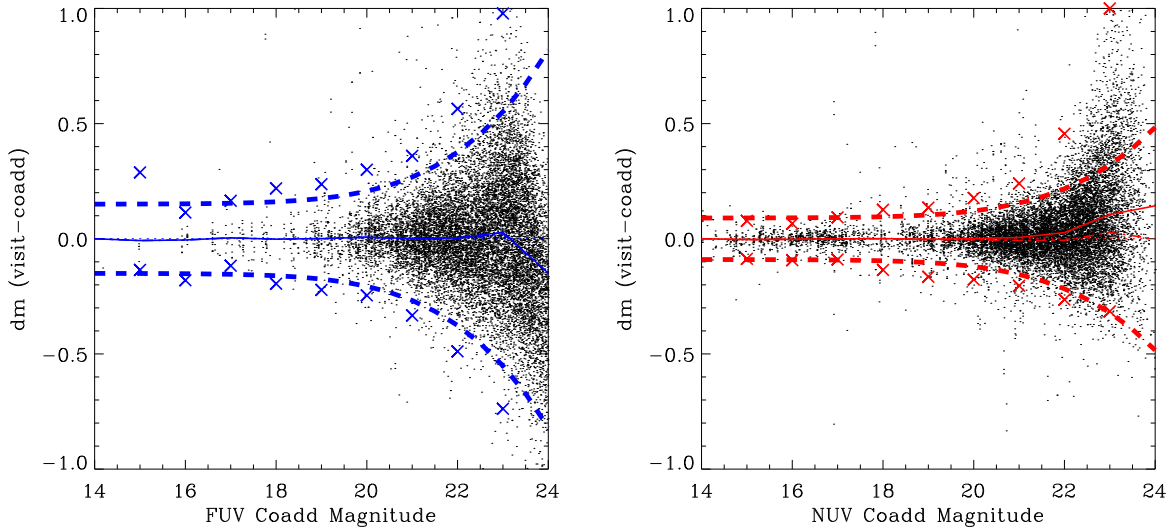}
\caption{GALEX photometric repeatability as a function of magnitude as
  determined by repeated measurements of sources in the GALEX deep fields (with
  individual measurements of order 1000~s).  In each panel, heavy dashed lines
  are the 3$\sigma$ values that would be expected given the flat field
  precision and Poisson counting statistics, while the $\times$ symbols show the
  actual value for each magnitude bin.  Solid overplotted lines show the median
  of the magnitude errors in each magnitude bin, while dot-dash lines show the
  same quantity after the visit magnitudes have been re-computed using the
  co-add background estimation.\label{dmvsm}}
\end{figure*}

The deep field \textit{visit} magnitude errors are also presented as a function
of the co-added ''truth'' magnitude in Fig.~\ref{dmvsm}, which serves to
illustrate an interesting bias effect on the faint end of the 1500~s depth MIS.
Here, the predicted $3\sigma$ errors (including the flat-field-limited values
derived for bright sources) are overplotted, showing good agreement with most
of the observed distribution but failing on the faint end.  This effect is
similar in magnitude but opposite in direction for each detector.  In the FUV,
sources are detected in the co-added image that are not detected in the visit
images except when the Poisson errors tilt the detection to the bright side
(the Eddington bias), thus faint FUV detections are biased bright by about
0.1~m$_{AB}$ compared to the ''truth'' catalog.  Similarly in the NUV, more
sources are detected in the co-added image, however the effect of the
undetected sources is to force an overestimate of the NUV background of about
5\% (with large scatter).  In this case, faint NUV detections are biased
\textit{dimmer} by about 0.1~m$_{AB}$ compared to the ''truth'' catalog.  In
each panel, solid overplotted lines show the median of the magnitude errors in
each magnitude bin, while dot-dash lines show the same quantity after the visit
magnitudes have been re-computed using the co-add background estimation.  This
correction is indistinguishable from the solid line median in the FUV panel
because the background is negligible, but completely cancels the faint end bias
in the NUV.  Since the effects of these two biases are in opposite directions
for each band, they have a more significant effect on the UV color, about
0.2~m$_{AB}$ at the faint end of the distribution.

\begin{figure*}
\includegraphics[width=3in,angle=90]{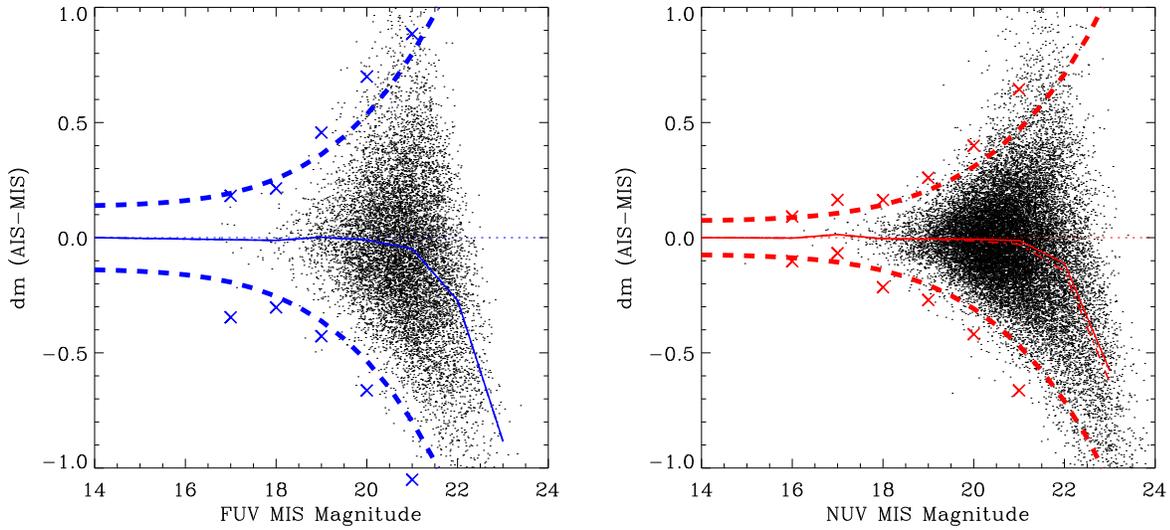}
\caption{GALEX AIS photometric repeatability as a function of magnitude as
  determined by repeated measurements of sources in overlapping MIS fields.  In
  each panel, heavy dashed lines are the 3$\sigma$ values that would be
  expected given the flat field precision and Poisson counting statistics,
  while the $\times$ symbols show the actual value for each magnitude bin.
  Solid overplotted lines show the median of the magnitude errors in each
  magnitude bin, while dot-dash lines show the same quantity after the visit
  magnitudes have been re-computed using the co-add background estimation (in
  the AIS this correction does not have a significant effect).  At the very
  faint end, there is an apparent bias toward visit magnitudes brighter than
  co-added magnitudes due to Eddington bias in the source-limited
  sample.\label{aisdmvsm}}
\end{figure*}

For completeness, we present a similar analysis of the photometric errors
in the 100~s AIS sample relative to cross-matched sources in overlapping MIS fields.  As
illustrated in Fig.~\ref{aisdmvsm}, we find that the AIS magnitudes are
biased bright at the faint end relative to the MIS for \textit{both} detectors,
and that corrections to the AIS magnitudes using MIS backgrounds have no effect
at all.  This result stems from the shallower depth of the AIS survey, which is
source rather than background-limited.

\subsection{Linearity}
There are two sources of photometric non-linearity in the instrument: global
non-linearity resulting from the finite period required for the electronics to
assemble photon lists (during which time new events are locked out), and local
non-linearity resulting from the MCP-limited current supply to small regions
around a bright sources.  The local effect complicates the NUV white dwarf
calibration measurements.  

Global effects are easily measured using the
on-board ``STIM'' pulser, which electronically stimulates each detector anode
with a steady, low rate stream of electronic pulses.  These are resolved as
point sources in the corners of the dose (\textit{-scdose}) images.  Since the real
rate of STIM pulses is accurately known, the measured rate is used by the
pipeline to scale the effective exposure.  The count rates for all sources in
the field are simultaneously corrected.  This effect is typically about 10\% in
NUV and a few percent in FUV, however it can become quite significant ($\sim
50$\%) for the brightest fields.

\begin{figure}
\includegraphics[width=2.25in,angle=90]{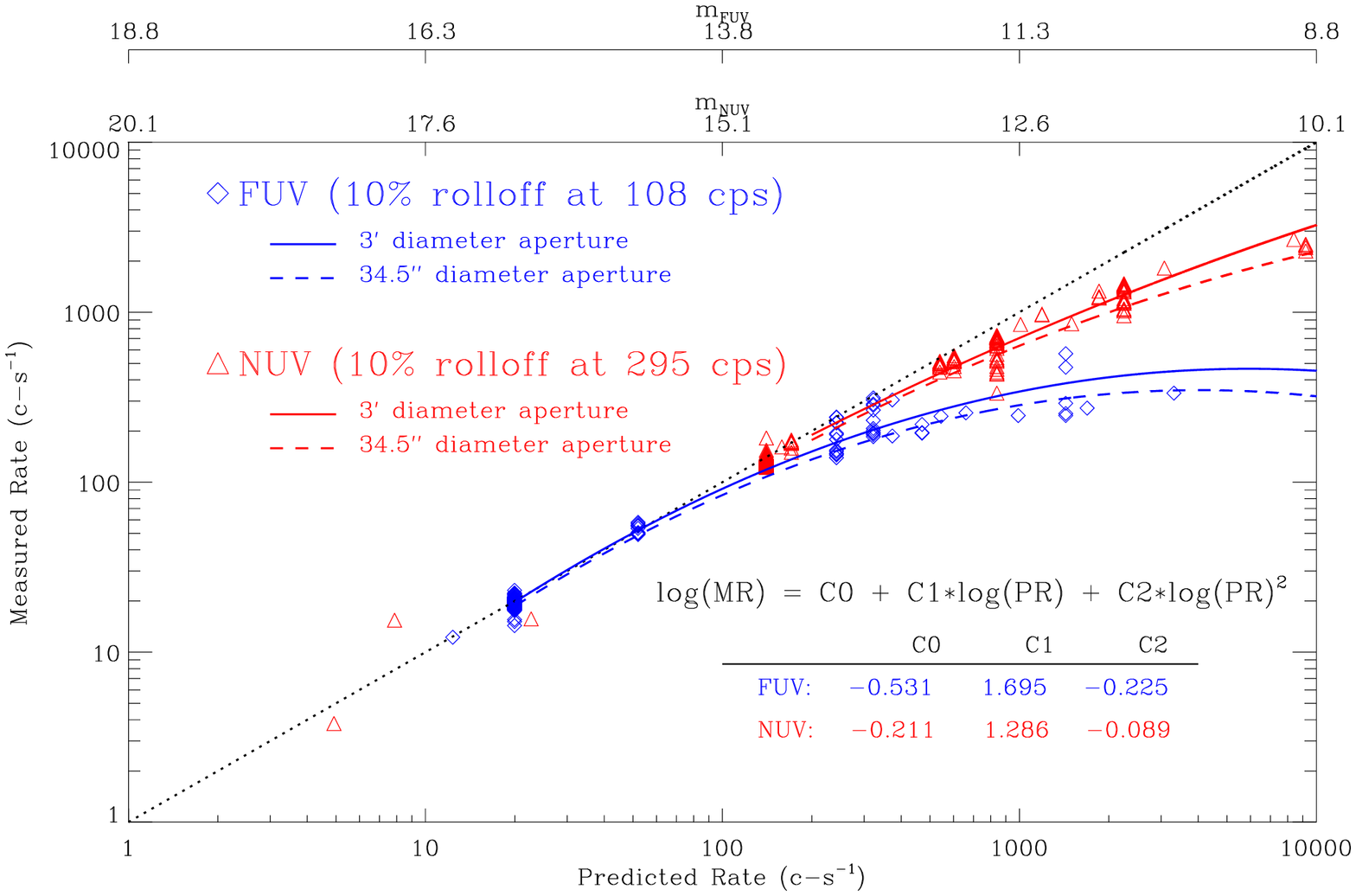}
\caption{Measured count rate versus predicted count rate for the white dwarf
  standards measured to date by GALEX using 34.5\arcsec\ (\textsc{aper\_7}) and
  3\arcmin\ diameter apertures.  These measurements represent a worst case
  because they are for point sources.  The non-linearity will be much
  \textit{worse} when measuring fluxes using small apertures, since this effect
  is concentrated at the core of each image.\label{locallin}}
\end{figure}

Local non-linearity can not be corrected well with the existing
calibration.  It affects not only the measured count rate of an individual
bright source but also the source shape and position.  Furthermore, the effects
vary around the field of view with the gain of the detector (as exemplified by
the white dwarf saturation correction described in Section~\ref{magref}).  We
have used standard stars to estimate the local non-linearity in each band as shown
in Fig.~\ref{locallin}.  This data may be used to approximately correct bright
sources that are within the measured range.  Note that the measurements
presented in the Figure are for 34.5\arcsec\ (\textsc{aper\_7}) and 3\arcmin\
diameter apertures; the effect will be much \textit{worse} when small apertures
are used.  The NUV detector is more robust to bright sources because its
photoelectrons are proximity-focused and thus present a larger source image
(with lower count density) to the MCP.  Local photometric non-linearity is
strongly source-size dependent. The saturation shown in Fig.~\ref{locallin} is
a worst-case scenario (using stars).  Wide scatter in measurements of bright
sources is an indication of the variation of the onset of saturation at
different detector locations.

We have increased the local count rate limit to 30,000~cps per source in
NUV, which corresponds to approximately 9th magnitude.  This new limit allows
most of the sky to be observed.  The FUV local limit is maintained at
5,000~counts-s$^{-1}$, which is not currently a survey constraint.  Global
detector limits are set in hardware (rather than the planning system) to
100,000~counts-s$^{-1}$, a limitation based largely on the boundaries of ground
testing and engineering judgment about safe rates of gas generation inside the
sealed detector tubes.  Areas such as the galactic plane and the Magellanic
Clouds are still restricted by the 100,000~cps global limit

\subsection{The GALEX Deep Fields}

The specific case of faint DIS photometry presents its own special challenges.  The relatively large GALEX PSF
affects the photometry of faint sources, mostly due to source confusion
\citep{xu2005}.  As an alternative to the GALEX pipeline for faint source
extraction, a Bayesian approach has been attempted by others
\citep{guillaume2006} to estimate the UV flux using prior information from
external, well-resolved images (mostly from visible wavelengths). The basic
idea is to generate an image with prior sources convolved with the GALEX PSF,
and to estimate what scaling factors to apply that will best fit the UV
image. The implemented method is based on a maximum likelihood parametric
optimization in the presence of Poisson noise. Scaling factors were extracted
via an iterative procedure to solve the set of non-linear equations based on
the Expectation-Maximization (EM) method.  In order to test the reliability of
the photometry extracted using this method, simulated objects (with a large
range of UV flux) were added to the GALEX DIS images. For an NUV DIS image with
60~ksec exposure, it was possible to recover the photometry of simulated sources
with an accuracy of 0.3~m$_{AB}$ to NUV$\sim$25.5, extending the faint-end
limit of the survey by about a magnitude.
 
\section{Resolution}
\label{resolution}
There are numerous contributors to the width of the GALEX PSF, including the
optics, detectors, and ground pipeline.  The reconstruction algorithms for
GR2/GR3 have improved to the point that the instrumental contributions are now
the largest share, particularly in the proximity-focused NUV channel.  As we
will show, the GALEX detectors dominate in this area and have significantly
position-dependent performance.

\begin{figure}
\includegraphics[width=3.25in]{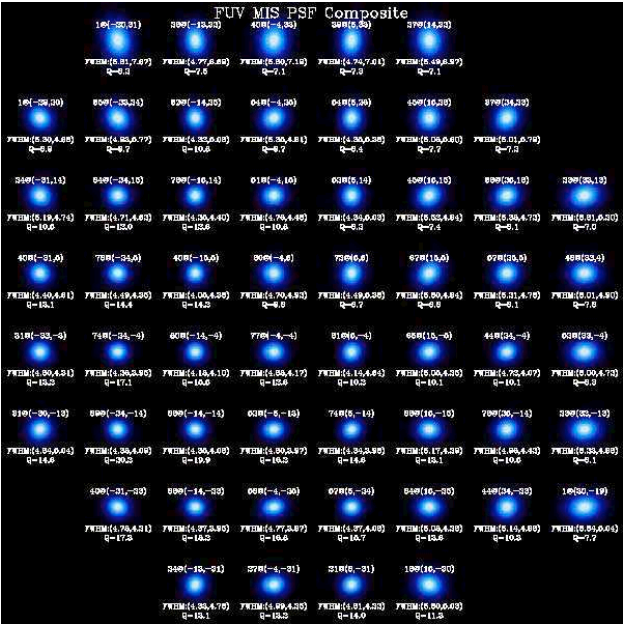}
\caption{Composite PSFs for the GALEX FUV channel in detector coordinates.  The
  individual images are exaggerated for clarity.
  Upper data values in the format N@(X,Y), where N is the number of stacked images and (X,Y)
  is the position in minutes of arc from the field center.  The FWHM is specified in
  seconds of arc for each detector axis.  Q is the typical pulse height at the
  specified detector location, in bins. The pulse height scale is not extremely well
  calibrated, but it is linear with 10~bins $\sim 1\times10^7$
  electrons).\label{composite_fuv_psf}}
\end{figure}

\begin{figure}
\includegraphics[width=3.25in]{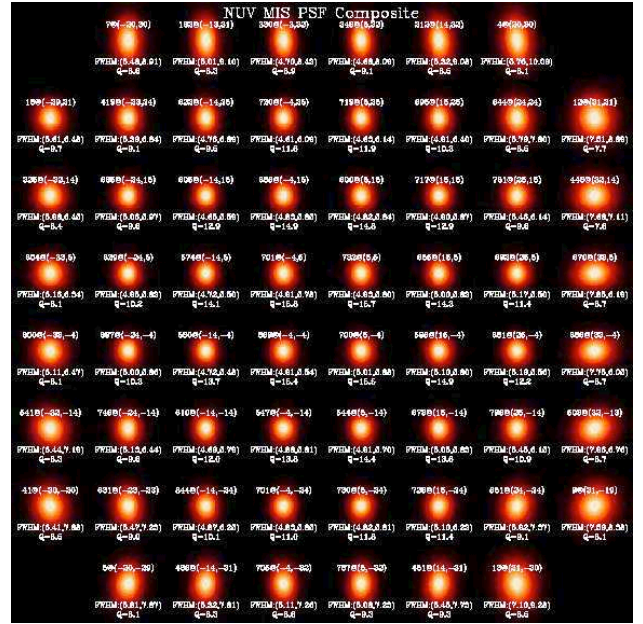}
\caption{Composite PSFs for the GALEX NUV channel in detector coordinates.  The
  individual images are exaggerated for clarity.  Upper data values are
  N@(X,Y), where N is the number of stacked images and (X,Y) is the position in
  minutes of arc from the field center.  The FWHM is specified in seconds of
  arc for each detector axis.  Q is the typical pulse height at the specified
  detector location, in bins.  The pulse height scale is not extremely well
  calibrated, but it is linear with 10~bins $\sim 1\times10^7$
  electrons).\label{composite_nuv_psf}}
\end{figure}

To verify the fundamental instrument performance from on-orbit data, some
bright stars were analyzed individually, outside the pipeline.  These results
show performance that is consistent with or better than what was measured
during ground tests.  We have also verified the end-to-end performance
including the pipeline by stacking images of stars from the MIS survey that
were observed at different locations on the detector.  The results of these
composites are shown in Figures~\ref{composite_fuv_psf}
and~\ref{composite_nuv_psf}.  Performance is reasonably uniform except at the
edge of the field, where it is significantly degraded.  Some remaining degraded
areas of the images may yet be improved with updates to the static calibration.  
Note that DIS surveys co-add data with many roll angles, so the DIS PSF is considerably more symmetrical across the field of view.

\begin{figure}
\includegraphics[width=2.5in,angle=90]{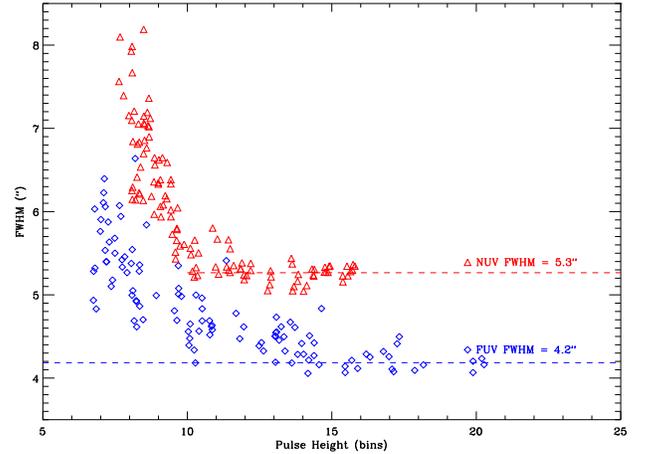}
\caption{The full width at half maximum of the GALEX PSF based on composite
  images for each detector as a function of \textit{intrinsic} detector pulse
  height.  In general, low pulse height regions are near the edge of the
  detector.\label{fwhm_vs_q}}
\end{figure}

Many of the regions with
degraded resolution correlate with low detector gain, an effect illustrated in
Fig.~\ref{fwhm_vs_q}.  The relative detector gain (or charge, 'Q') in the
region of measurement for each source is tabulated in the merged catalog.  This
gain measurement is obtained from a look-up table that represents the detector
in an unsaturated condition.  The \textit{actual} gain measured during the
observation is available in the pulse height (\textit{-scq}) image data product.

\begin{figure}
\includegraphics[width=2.5in,angle=90]{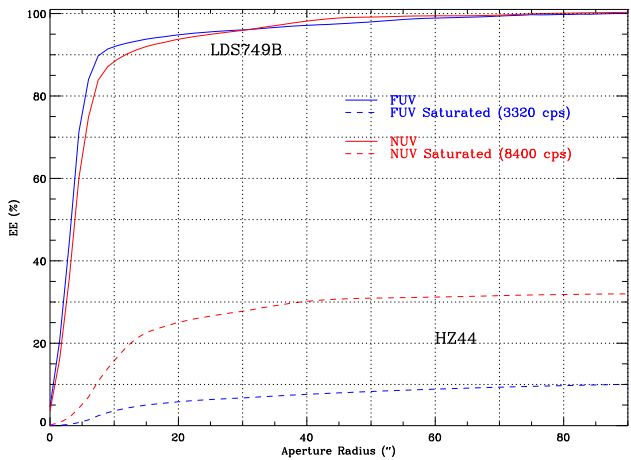}
\caption{GALEX FUV and NUV composite encircled energy curves for two white
  dwarf standards.  A composite of LDS749b measurements forms the unsaturated
  (solid) curve, while the brightest standard that we have observed, HZ~44,
  defines the saturated curve.  The reference count rates for the saturated
  curves are the predicted values; measured rates are much
  lower as indicated by the low encircled energy values.\label{extended_ee_curve}}
\end{figure}

It is also evident that the GALEX PSF has structure a significant distance from
the center.  We have characterized this by building a composite PSF of the
white dwarf calibrator LDS749b as shown in Fig.~\ref{extended_ee_curve}.  The
effects of local non-linearity are also illustrated here by the highly
saturated white dwarf HZ~44, the brightest we have observed with GALEX in
imaging mode.  The effect of the saturation is strongest in the core of the
PSF, causing the wings to be strongly emphasized.

\section{Astrometry}
\label{astrometry}
The GALEX GR2/GR3 astrometric precision represents an improvement of order 20\%
over GR1 (and much more compared to the Early Data Release).  The dynamic
pipeline transformation is generally not a limiting factor as it had been in
earlier releases.  The static transformation (primarily the distortion map) is
also performing very well for most of the field of view.  Improvements could
still be made that would not only improve the global astrometric precision but
also the point spread function (PSF) at the edge of the field.

The distortion maps for the GR2/GR3 calibration are the result of several
flight-based iterations on the pre-flight calibration.  The first, which was
implemented for the GR1 data release, used Tycho-2 catalog \citep{hog2000}
reference stars to refine the distortion map for each detector independently.
An iterative technique was used until the measured GALEX positions converged
with the reference positions, a technique that relied on the assumption that
the distortion across a $\sim1$\arcmin\ ($\sim 1$~mm) region of the detector
(the footprint of the dither pattern) is reasonably represented by a single
average value.  This technique worked well, however there were relatively few
FUV stars available to perform the analysis (a few thousand).  The FUV-NUV
relative astrometry retained some noticeable ($\sim 1$\arcsec) misalignments.
We have addressed this issue with cross-matched SDSS data.  Since the SDSS
astrometry is better than $0.1$\arcsec\ RMS \citep{pier2003}, it can be
regarded as essentially perfect for our purposes.  The sample we gathered
contains 69642 objects spectroscopically classified as quasars (SDSS
\textsc{specClass} = 3, a convenient point source sample compiled for other
applications).  Objects were considered matches if they fell within 2.5\arcsec\
of each other.  There was no discrimination based on location in the GALEX
field of view.  We binned these results into a relatively coarse grid ($\sim
1$\arcmin), the intent of which was to ``stretch'' the existing plate solution
to remove low spatial frequency errors.  The result of this improvement is that
the GR2/GR3 absolute astrometric precision achieved by the GALEX MIS within
$0.6^{\circ}$ of the field center for bright (signal-to-noise ratio $\ge 10$)
point sources is 0.49\arcsec\ RMS in the NUV and 0.59\arcsec\ RMS in the FUV,
with similar relative precision between the two bands.  

\begin{figure}
\includegraphics[width=2.25in,angle=90]{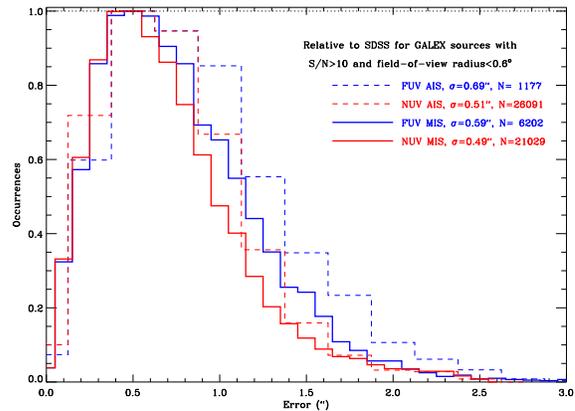}
\caption{Distribution of absolute astrometric errors in the
  GALEX GR2/GR3 survey data for quasars with S/N$ \ge 10$ in each band
  matched against the SDSS.\label{g2_astrometry}}
\end{figure}

\begin{figure}
\includegraphics[width=2.25in,angle=90]{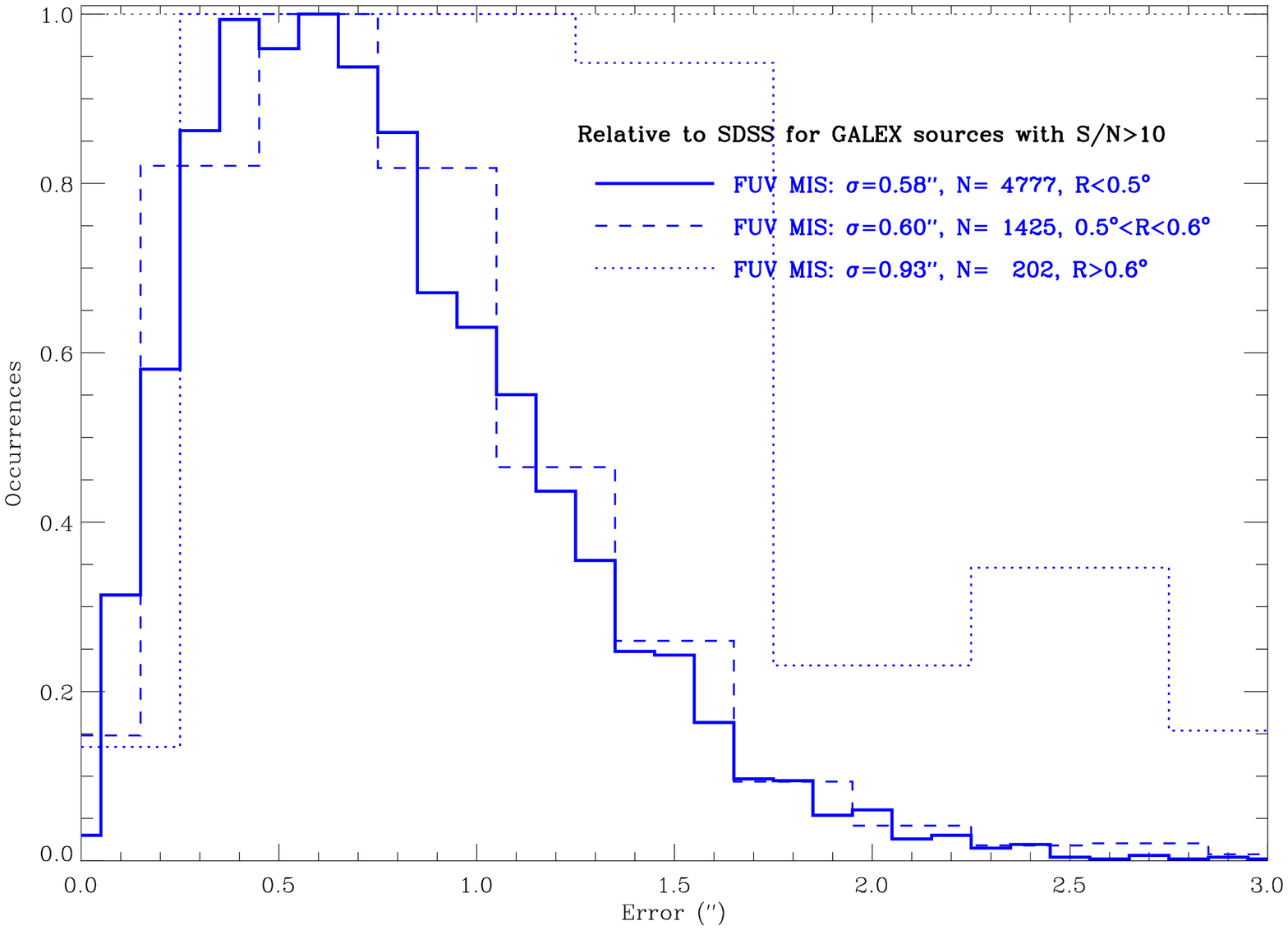}
\caption{FUV GR2/GR3 MIS absolute astrometric performance for field center and
  field edge distributions of SDSS quasars with GALEX
  S/N$ \ge 10$\label{radialposfuv}}
\end{figure}

\begin{figure}
\includegraphics[width=2.25in,angle=90]{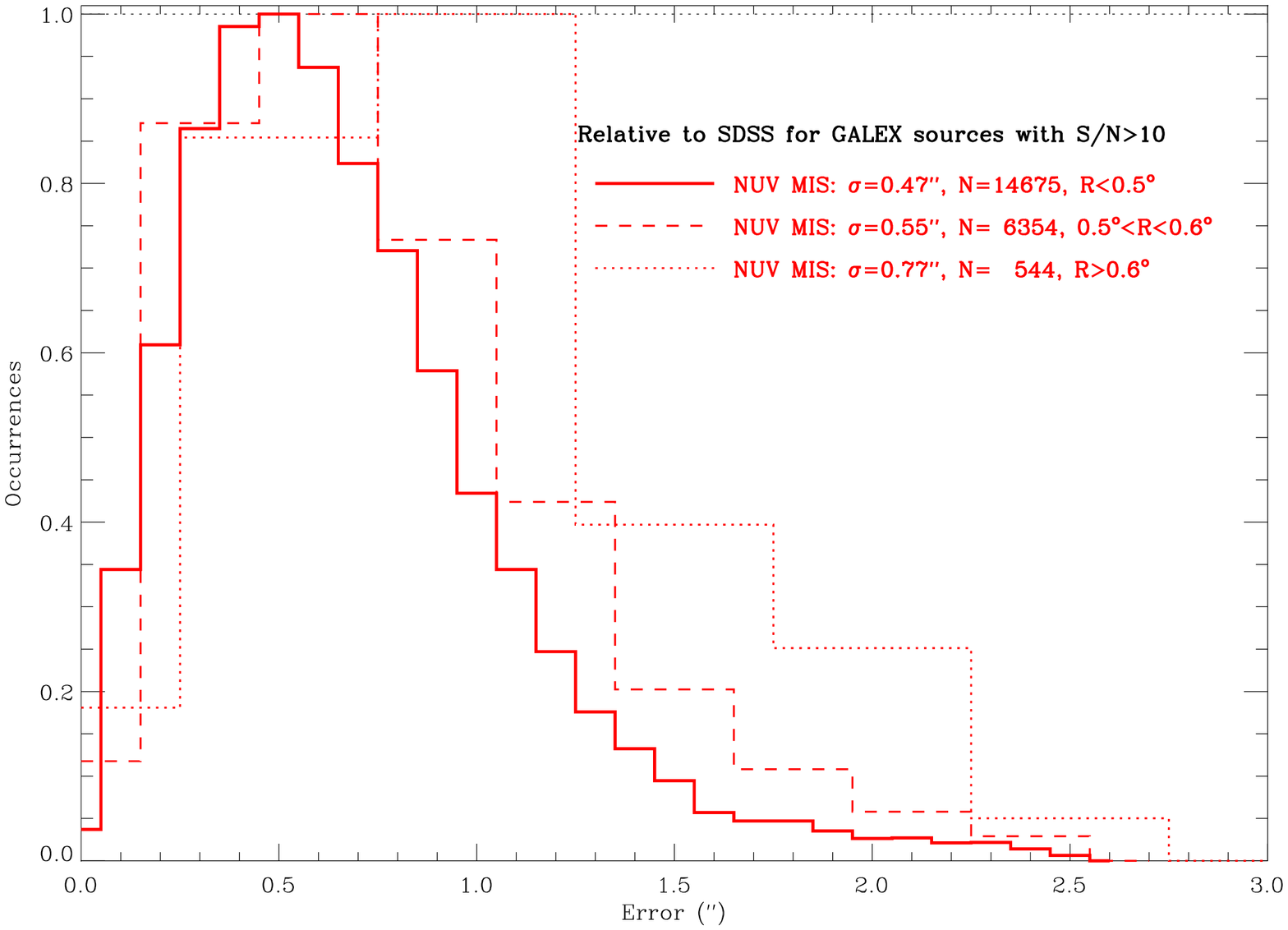}
\caption{NUV GR2/GR3 MIS absolute astrometric performance for field center and
  field edge distributions of SDSS quasars with GALEX S/N$
  \ge 10$\label{radialposnuv}}
\end{figure}

\begin{figure}
\includegraphics[width=2.25in,angle=90]{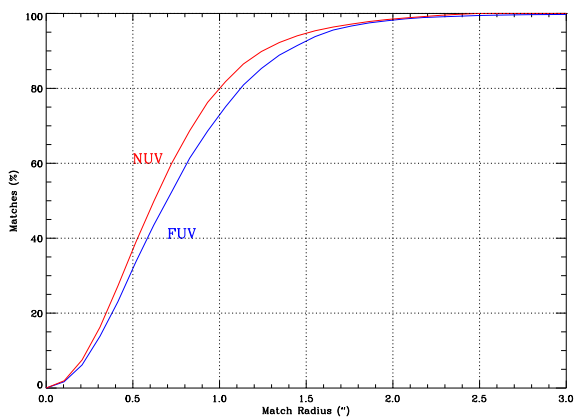}
\caption{The percentage of GALEX-SDSS matches as a function of matching radius.
  An MIS sample with S/N$ \ge 10$ and field radius less than $0.6^{\circ}$ was
  used for this analysis.  Note that the matched sample is limited to
  2.5\arcsec\ maximum error, although based on the shape of the distribution
  this does not appear to be an issue.\label{matchpercent}}
\end{figure}

In Fig.~\ref{g2_astrometry} we present histograms of the position errors for
high S/N SDSS quasars detected by GALEX.  The most noticeable image defect
resulting from the current astrometric solution is the degraded PSF at the edge
of the field.  As shown in Figures~\ref{radialposfuv}
and~\ref{radialposnuv}, the astrometric precision in the outer 5\arcmin\ of the
detector, while degraded, still performs nearly at the level of the central
30\arcmin.  Generally speaking, matches between GALEX and other surveys can
be safely performed with a 2\arcsec\ window as indicated by
Fig.~\ref{matchpercent}.  

The most promising future improvement to the astrometric calibration probably
involves a new pipeline module that would scale the amplitude of the distortion
maps by a factor determined from measurements of the STIM pulses at the corners
of each image.  Much of the necessary data to eliminate this ``breathing'' mode
is already available.  Another more complex possibility involves reprocessing
the reference star data in $\sim20$~s intervals to reduce the footprint of each
source sample on the detector and thus more finely sample the detector grid.
Such an analysis would be most useful near the edge of the detector where the
distortion map is changing rapidly.

\section{Spectroscopy}
\label{spectroscopy}
\begin{figure}
\begin{center}
\includegraphics[width=3in,angle=270]{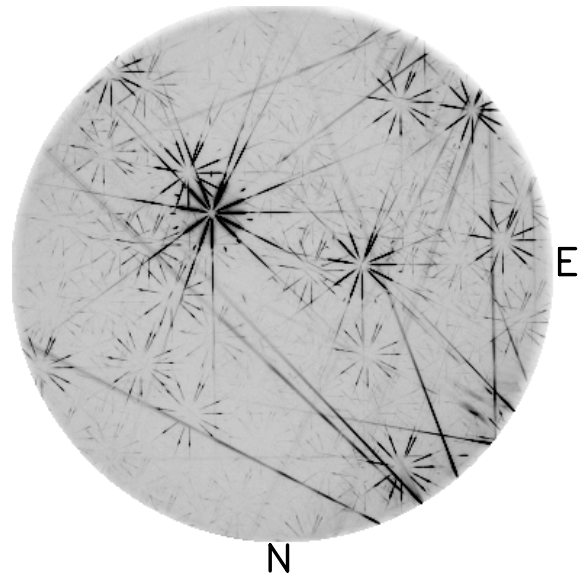}
\caption{A composite image of GALEX grism spectra collected with multiple grism
   position angles.  As the grism is rotated in its mount, each spectrum spins
   around its own undeviated wavelength point to form a ``pinwheel'' in the
   composite, illustrating the optimal grism angle (producing an unconfused
   spectrum) for each source.\label{grismpinwheel}}
\end{center}
\end{figure}

The spectroscopic mode utilizes a CaF$_{\rm 2}$ grism (a prism with a ruled
surface) that can be moved into the converging beam of the telescope to form
simultaneous (slitless) spectra of all sources in the field in both bands.
Many spectra will normally overlap, a problem addressed by rotating the grism
on its axis to different position angles during repeat observations of the same
field.  Each spectrum has an ``undeviated wavelength point'' around which it
will rotate as the grism position angle is changed (thus any given spectrum
will have an optimal set of grism position angles depending on the location of
nearby sources).  As the position angle changes, the undeviated wavelength
point remains fixed in the field as illustrated by Fig.~\ref{grismpinwheel}.
The resulting spectra can be combined to eliminate confusion.
The standard GALEX spectroscopy pipeline is designed for isolated point sources,
which are detected in the GALEX direct image observations.  Extended or
overlapping sources, or sources which do not appear in the direct image catalog
require a special pipeline reduction with modified parameters.

A relative position correction as a function of time is applied to the spectral
image data using a technique very similar to that used for direct image
photons.  Using the brightest sources from the catalog, an aspect
correction is applied to the entire exposure by fitting four parameters: shift
in dispersion direction, shift in spatial direction, spacecraft roll angle, and
grism position angle.  A spectrum points North (blue to red) when the grism
position angle is $0^{\circ}$ and towards the East when it is $90^{\circ}$.

Bright sources may show grism orders ranging from -4 to +5 across the field,
however only two orders in each band are typically extracted.  For NUV, the
highest response is in first order with a total system peak response of
40~cm$^2$.  For FUV, the strongest order is 2nd with a peak response of
20~cm$^2$.  The dispersion angles $\gamma_{FUV}$ and $\gamma_{NUV}$ in
arcseconds from the undeviated wavelength point for FUV 2nd order and NUV 1st
order are given by:

\begin{tiny}
\begin{eqnarray*}
\gamma_{FUV} & = & -4530.7 + 7.3859\lambda - 3.9254\times 10^{-3}\lambda^2 + 7.4750\times 10^{-7}\lambda^3 \\
\gamma_{NUV} & = &  -882.1 + 7.936\times 10^{-1}\lambda - 2.038\times 10^{-4}\lambda^2 + 2.456\times 10^{-8}\lambda^3
\end{eqnarray*}
\end{tiny}
where $\lambda$ is the wavelength in Angstroms, accurate to a few tenths of an arcsecond.  

\begin{figure*}
\includegraphics[width=6in,angle=270]{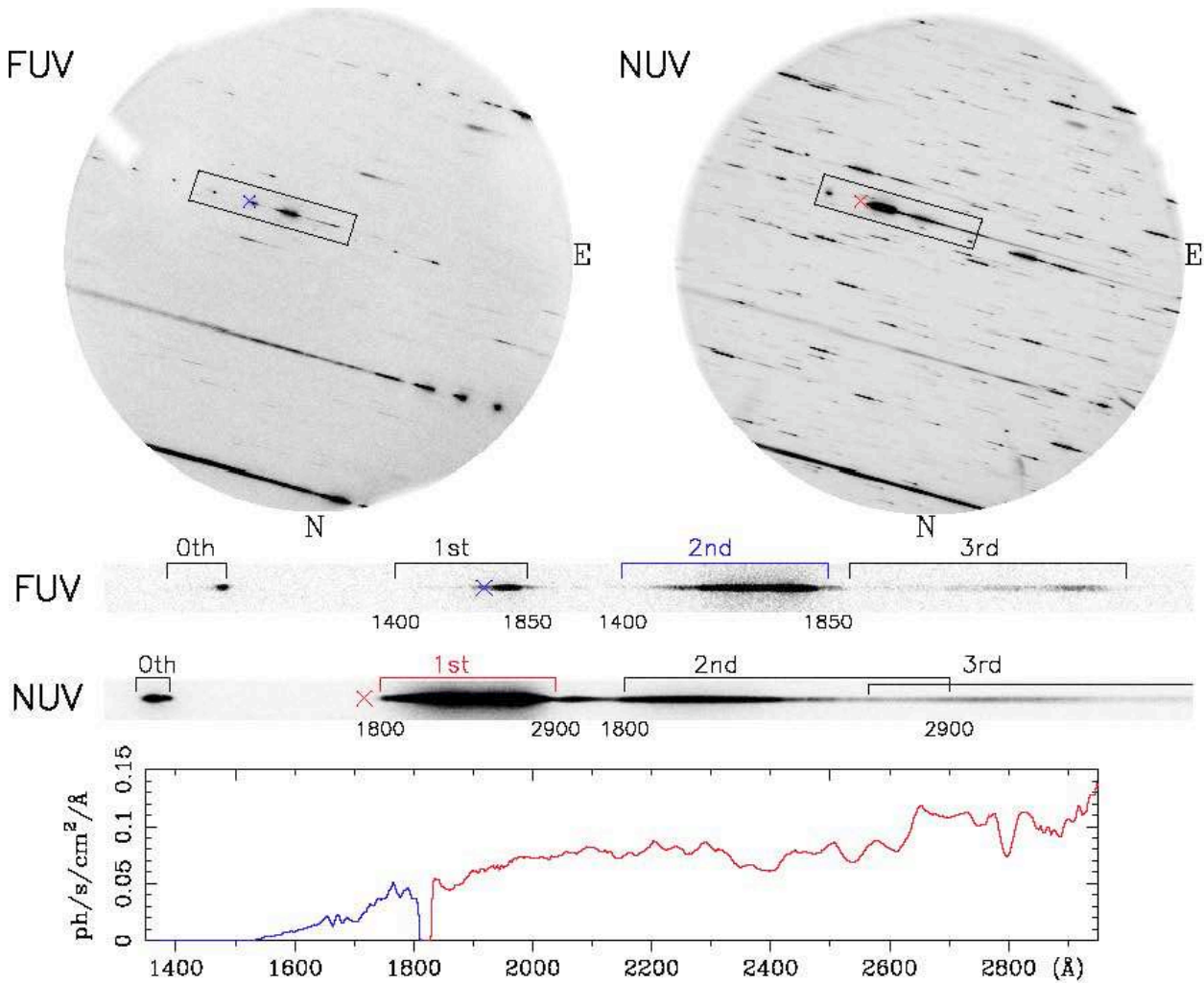}
\caption{Example bright object GALEX spectra.  Circular images at top show the
  reconstructed FUV and NUV spectral image data.  A rectangular region of
  interest identifies the same spectrum in each band, while an $\times$ marks the
  undeviated wavelength point around which the spectra would spin if the grism
  were rotated.  The rectangular regions are enlarged and annotated in the
  middle of the frame, where wavelength ranges and spectral orders are
  identified.  In the bottom panel, the calibrated spectrum is shown in blue
  for the FUV portion and red for the NUV portion.\label{grism}}
\end{figure*}

Using the catalog positions, image strips (aligned to the dispersion direction)
are created for each source using the corrected photon positions.
Fig.~\ref{grism} shows how a portion of the field image translates into a
dispersion-aligned image strip for each band.  The Figure shows image strips
covering 1700 arcseconds in the dispersion direction to illustrate the weaker
orders.  Only 900 arcseconds (from -100 to +800 arcseconds relative to the
undeviated point) covering the two strongest orders in each band are stored
during actual data reductions.  The width of each image strip is 78 arcseconds.
Fig.~\ref{grism} also shows the undeviated point in each image strip (marked
with an $\times$) which occurs at about 1700~\AA\ in 1st order in both bands.
This is the point where the spectral images for all position angles intersect.

The spectral shape, strength, and position of neighboring sources are estimated
from the direct image catalog and are masked from each image strip.  Additional
masking is done on each strip when the exposures are coadded to eliminate other
artifacts not predicted by known neighboring sources, which may be due to
reflections, ghosts, hot spots, or unidentified sources near the edge of the
field.  The response map is then applied and the background is fit and
subtracted at each wavelength point.  An aperture extraction (summation) is
computed for the primary spectrum using the brightest order in each band (NUV
1st and FUV 2nd).  An optimal extraction is also computed using a PSF model.  A
20 Angstrom region is set to zero in the spectrum at around 1820~\AA\
indicating the separation between FUV and NUV.

A list of sources for which spectra will be extracted is determined
  by estimating what the final S/N of a given source will be after coadding all grism
  observations.
  Any source which will achieve a S/N per resolution element in
the coadded result of $\geq 2$ in FUV or $\geq 3$ in NUV is selected for
spectral extraction.  
For multi-exposure data reductions, the masked image-strip photon
  and response data are coadded for each source.
  For both
individual and coadded spectra, a response correction is applied as a function
of wavelength and the spectra are rebinned to a linear scale.  The primary (NUV
1st order and FUV 2nd order), secondary (NUV 2nd order and FUV 3rd order), and
optimal (the primary spectrum with PSF weighting in the spatial direction)
extractions are stored in 3 arrays in the final reduced data file.  The
estimated S/N for unresolved sources in single-orbit spectra is shown in
Tab.~\ref{specsn}.
\begin{deluxetable}{ccc}
\tabletypesize{\scriptsize}
\tablecaption{Typical signal-to-noise ratio for unresolved GALEX spectra at MIS
  depth (1500~s).\label{specsn}}
\tablewidth{0pt}
\tablehead{\colhead{m$_{AB}$} & \colhead{FUV S/N} & \colhead{NUV S/N}}
\startdata
16       & 10.4    & 21.2\\
18       & 3.4     & 6.1\\
20       & 0.79    & 1.22\\
22       & 0.14    & 0.20\\ 
\enddata
\end{deluxetable}

The usable wavelength range is approximately 1300 to 1820~\AA\ for FUV (2nd order) and
1820 to 3000~\AA\ for NUV (1st order).  
The average dispersion is 1.64 and 4.04 \AA\ per arcsecond for FUV (2nd order)
  and NUV (1st order), respectively.
The average resolution ($\lambda/\Delta\lambda$) 
for a point source (assuming a 5 arcsecond PSF) is about 200 (8\AA) and 118 (20\AA)
for FUV (2nd order) and NUV (1st order), respectively.

Although there is evidence that the response may vary with position on the
detector or with the grism position angle, a single wavelength-dependent
calibration is used for each band.  These variations may introduce a
calibration error of about 3\% in the body of each spectral order and about
10\% near the edges of each order.

\section{Conclusions}
\label{conclusions}
We have described the GALEX performance and calibration results for the GR2 and
GR3 data releases.  These releases have identical pipeline calibrations that
are significantly improved over the previous GR1 release.  The analysis we
have presented provides an overview of the data quality, errors and
peculiarities that are likely to confront astronomers.

The relative photometric precision is 0.05 and 0.03 m$_{AB}$ in the FUV and NUV
respectively.  We have identified several photometry issues that will be
improved in future calibrations.  The first is a systematic drift with time of
order 1\% in FUV and 6\% in NUV, total, over the period of the released data
($\sim$4 years).  The sense of the drift is that objects appear to be growing
fainter with time.  The second issue is that the current FUV calibration relies
on data collected in a non-standard mode during the 2005 FUV recovery, and this
data appears bright by 5\% compared to other measurements of the same objects
collected in the standard mode.  The third issue is that the NUV zero point is
defined by a white dwarf calibrator that is at least partially saturated.
While we have evidence that the saturation is mild and correctable, we are
planning confidence-building observations of an additional white dwarf, the
dimmest in the CALSPEC catalog, to help resolve the issue.

The astrometric precision has been improved, about 20\%, to
0.5\arcsec\ RMS.  There is evidence that it may be improved further by
accounting for the detector ''breathing'' mode already measured with existing
data from an external pulser that runs in parallel during all detector
operations.

All of our performance parameters continue to meet flight requirements.  As the
mission continues, we naturally expect the calibration to evolve.  The
instrument itself continues to function nominally in spite of a series of
flight anomalies especially with the FUV detector.  We have modified operations
and flight software over the mission to respond to these issues. While some of
them have significantly affected the FUV observation efficiency, none give any
indication of affecting the instrument performance.  We believe there is room
to increase the count rate limits for both detectors (especially FUV), and
to the extent possible we plan to explore higher rate fields closer to
the Galactic plane after the primary surveys are completed.

\acknowledgments

GALEX (Galaxy Evolution Explorer) is a NASA Small Explorer, launched in April
2003.  We gratefully acknowledge NASA's support for construction, operation,
and science analysis for the GALEX mission, developed in cooperation with the
Centre National d'Etudes Spatiales of France and the Korean Ministry of Science
and Technology.

{\it Facilities:} \facility{GALEX}

\clearpage
\begin{turnpage}
\begin{deluxetable*}{llcccl}
\tabletypesize{\footnotesize}
\tablewidth{0pt}
\tablecaption{A summary of GALEX pipeline product array content and format.\label{arrayproducts}}
\tablehead{
\colhead{Product} & \colhead{Extension} & \colhead{Format}  & \colhead{Units} & \colhead{Scale} & \colhead{Description}}
\startdata
Artifact Flags        & -flags    & $480\times480$   &                 &
12\arcsec-pixel$^{-1}$  
   & Artifact flags for each pixel\\
Background-Subtracted Intensity & -intbgsub 
   & $3840\times3840$ & counts-s$^{-1}$ & 1.5\arcsec-pixel$^{-1}$ 
   & Intensity-Sky Background\\
Count                 & -cnt      & $3840\times3840$ & counts          &
1.5\arcsec-pixel$^{-1}$ 
   & Aspect-corrected image, no flat field.\\
Dose                  & -scdose   & $2250\times2250$ & counts          &
3\arcsec-pixel$^{-1}$   
   & Detector space, corrected for walk, wiggle and distortion.\\
Exposure              & -exp      & $960\times960$   & s               &
6\arcsec-pixel$^{-1}$   
   & Exposure, \textit{un}corrected for flat field or dead time \\
Intensity             & -int      & $3840\times3840$ & counts-s$^{-1}$ &
1.5\arcsec-pixel$^{-1}$ 
   & Count image divided by relative response image\\
Movie                 & -movie    & $480\times480$   & counts          &
12\arcsec-pixel$^{-1}$  
   & Count map slices at 16~s intervals\\
Object Mask           & -objmask  & $3840\times3840$ &                 &
1.5\arcsec-pixel$^{-1}$ 
   & Areas masked during sky background estimation\\
Pulse height          & -scq      & $2250\times2250$   & bins          &
3\arcsec-pixel$^{-1}$   
   & Average pulse height per pixel\\
Relative response     & -rrhr     & $3840\times3840$ & s               &
1.5\arcsec-pixel$^{-1}$ 
   & Effective exposure, corrected for dither, flat field and dead time \\
Relative response     & -rr       & $960\times960$   & s               &
6\arcsec-pixel$^{-1}$   
   & Effective exposure, corrected for dither, flat field and dead time \\
Sky Background        & -skybg    & $3840\times3840$ & counts-s$^{-1}$ &
1.5\arcsec-pixel$^{-1}$ 
   & Sky background estimate\\
Threshold             & -wt       & $3840\times3840$ &                 
   & 1.5\arcsec-pixel$^{-1}$ & Background-subtracted intensity map divided by
   estimated Poisson noise\\
\enddata
\end{deluxetable*}
\end{turnpage}

\end{document}